\documentclass[%
 aip,
 amsmath,amssymb,
 reprint,%
]{revtex4-1}

\usepackage{graphicx}% Include figure files
\usepackage{dcolumn}% Align table columns on decimal point
\usepackage{bm}% bold math
\usepackage[utf8]{inputenc}
\usepackage[T1]{fontenc}
\usepackage{mathptmx}
\usepackage{etoolbox}
\usepackage{comment}
\usepackage{xcolor}

\makeatletter
\def\@email#1#2{%
 \endgroup
 \patchcmd{\titleblock@produce}
  {\frontmatter@RRAPformat}
  {\frontmatter@RRAPformat{\produce@RRAP{*#1\href{mailto:#2}{#2}}}\frontmatter@RRAPformat}
  {}{}
}%
\makeatother
\begin{document}

\preprint{AIP/123-QED}

% New title
\title{Harmonically Induced Shape Morphing of Bistable Buckled Beam with Static Bias\\}

\author{Md Nahid Hasan}
\affiliation{Department of Mechanical Engineering, University of Utah, Salt Lake City, UT 84112, USA}
\affiliation{Department of Mechanical Engineering, Montana Technological University, Butte, MT 59701, USA}

\author{Sharat Paul}%
\affiliation{Department of Mechanical Engineering, University of Utah, Salt Lake City, UT 84112, USA}

\author{Taylor E. Greenwood}%
\affiliation{Department of Mechanical Engineering, University of Utah, Salt Lake City, UT 84112, USA}
\affiliation{Department of Mechanical Engineering, Pennsylvania State University, University Park, PA 16802, USA}

\author{Robert G. Parker}%
\affiliation{Department of Mechanical Engineering, University of Utah, Salt Lake City, UT 84112, USA}

\author{Yong Lin Kong}%
\affiliation{Department of Mechanical Engineering, University of Utah, Salt Lake City, UT 84112, USA}
\affiliation{Department of Mechanical Engineering, Rice University, Houston, TX 77005, USA}
%\thanks{Authors to whom correspondence should be addressed: yong.kong@utah.edu}

\author{Pai Wang}%
\affiliation{Department of Mechanical Engineering, University of Utah, Salt Lake City, UT 84112, USA}
\thanks{Authors to whom correspondence should be addressed: pai.wang@utah.edu}

\date{\today}% It is always \today, today,
             %  but any date may be explicitly specified
\begin{abstract}
We investigate the effect of a constant static bias force on the dynamically induced shape morphing of a pre-buckled bistable beam, focusing on the beam's ability to change its vibration to be near different stable states under harmonic excitation. Our study explores four categories of oscillatory motions: switching, reverting, vacillating, and intra-well in the parameter space. We aim to achieve transitions between stable states of the pre-buckled bistable beam with minimal excitation amplitude. Our findings demonstrate the synergistic effects between dynamic excitation and static bias force, showing a broadening of the non-fractal region for switching behavior (i.e., switching from the first stable state to the second stable state) in the parameter space. This study advances the understanding of the dynamics of key structural components for multi-stable mechanical metamaterials, offering new possibilities for novel designs in adaptive applications.
\end{abstract}
\maketitle
%\keywords{Suggested keywords}%Use showkeys class option if keyword                         %display desired
%\textit{Introduction:}
Bistable buckled beams find application across mechanical metamaterials~\cite{hasan2023optimization,shan2015multistable,mei2021mechanical,gupta2024evidence,de2024nonlinear}, energy harvesters~\cite{ando2014bistable,harne2013review,ibrahim2017dynamics,harne2017harnessing,yang2020dynamics}, programmable mechanical devices~\cite{chen2021reprogrammable,meng2022deployable}, energy absorbers~\cite{restrepo2015phase,paul2024effects}, and MEMS devices~\cite{cao2021bistable,qiu2004curved,hasan2018novel}. Bistable systems, known for their two stable equilibrium states, offer a promising avenue for applications that require morphing~\cite{pontecorvo2013bistable,chi2022bistable} and reconfiguration~\cite{librandi2021programming,librandi2020snapping,zhang2017exploiting,wang2023phase,li2021reconfiguration}. However, quasi-static shape morphing of bistable buckled beams is known to be energy-intensive and time-consuming~\cite{yan2019analytical,camescasse2013bistable,cleary2015modeling,liang2023programmable,ten2024single,ghavidelnia2023curly}. 
%Vibration-induced shape morphing offers an alternative, requiring lower actuation amplitude than quasi-static methods and benefiting from the nonlinear dynamics of bistable systems~\cite{bonthron2024dynamic,PhysRevE.108.L022201,rouleau2024numerical}.
Vibration-induced shape morphing offers an alternative to quasi-static methods, requiring lower actuation amplitude and leveraging the inherent nonlinear dynamics of bistable systems~\cite{bonthron2024dynamic,PhysRevE.108.L022201,rouleau2024numerical}.
%Utilizing dynamic excitation effectively facilitates switching, reverting, and vacillating behavior, thereby achieving efficient shape morphing of the bistable beam.
By utilizing dynamic excitation, vibration-induced shape morphing effectively facilitates switching, reverting, and vacillating behavior, resulting in rapid and energy-efficient shape morphing~\cite{PhysRevE.108.L022201}. Nonetheless, the parameter space of a symmetric bistable system reveals that switching and reverting behaviors coexist in an intertwined chaotic region~\cite{PhysRevE.108.L022201,PhysRevE.57.1544,amor2023nonlinear,virgin2000introduction,moon2008chaotic,moon1984fractal,xu2024chaotic}. This coexistence creates challenges in selecting the appropriate combination of forcing amplitude and frequency to switch between stable states.
%It poses challenges in choosing an appropriate combination of forcing amplitude and frequency from the parameter space to switch between stable states efficiently. 
%Exploiting asymmetric bistability~\cite{kovacic2008resonance,sadeghi2020dynamic,simo2016effects} and broadening the switching behavior region within the parameter space of a bistable buckled beam can enhance the feasibility of using dynamic excitation for shape morphing of the buckled beam. 
By exploiting asymmetric bistability~\cite{kovacic2008resonance,sadeghi2020dynamic,simo2016effects}, the switching behavior region within the parameter space of a bistable buckled beam can be broadened, thereby enhancing the feasibility of using dynamic excitation for shape morphing. Recent research demonstrates that we can tune the quasi-static response of magnetized buckled beams under applied magnetic fields—specifically, their quasi-static force-displacement curves and energy landscapes exhibit asymmetric bistable behavior \cite{abbasi2023snap,pal2023programmable,zhao2023tuning,zhao2023encoding}. The translation of the force-displacement curve up or down 
along the force axis is linearly proportional to the applied magnetic field \cite{abbasi2023snap,pal2023programmable}.

In this letter, we apply a static bias force in combination with dynamic excitation, exploiting asymmetric bistability to morph the bistable buckled beam between stable states. Our investigation focuses on enhancing the morphing of bistable beams through the synergistic combination of static bias forces and low-amplitude dynamic excitation to switch between the states of the bistable buckled beam. By manipulating the energy landscape using a static bias force, we aim to achieve faster and more reliable control over the bistable buckled beam's dynamic transitions, thereby broadening the switching behavior region in the parameter space of a bistable buckled beam and increasing the predictability of shape morphing. Our approach advances the theoretical understanding of bistable beam dynamics and opens new avenues for efficient shape morphing of bistable systems.
\begin{figure}[htb!]
\centering
\includegraphics[width=\linewidth]{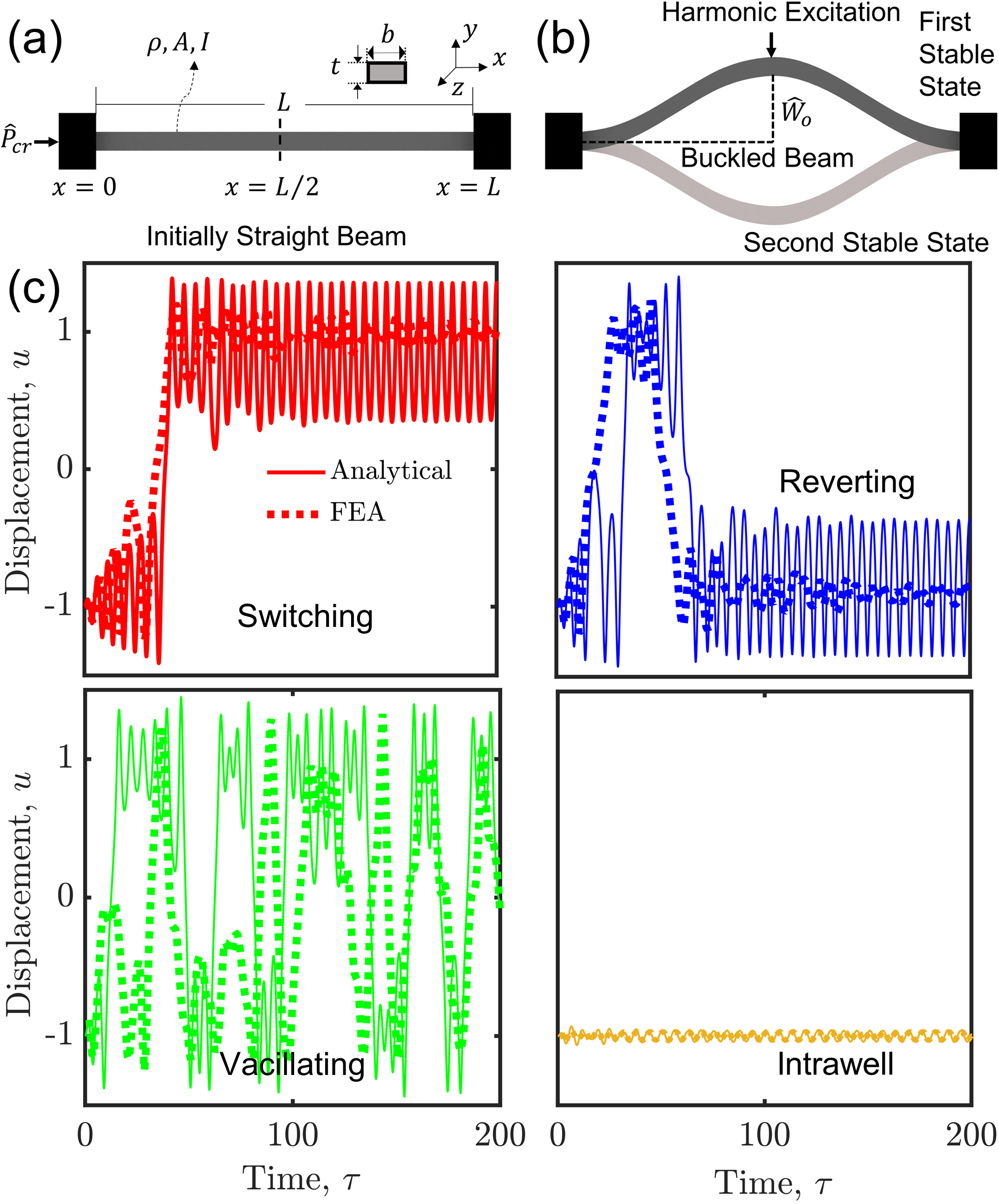}
\caption{Shape morphing of a bistable buckled beam exploiting symmetric bistability and dynamic excitation: (a) Schematic diagram of the beam's initial straight configuration. (b) Initially, the straight beam is compressed past the critical buckling load, \(\hat P_{\text{cr}}\), to achieve a static deflection denoted as \(\hat{W}_0\) with the fixed end. It can subsequently be dynamically excited to switch between its two stable states. (c) Analytical and finite element analysis (FEA) results showing switching behavior with dimensionless parameters \((G,\Omega)=(0.100,1.17595)\), reverting behavior with parameters \((G,\Omega)=(0.125,1.209412)\), aperiodic vacillating behavior with parameters \((G,\Omega)=(0.175,1.13291)\), and intra-well behavior with parameters \((G,\Omega)=(0.150,1.30506)\), all under a constant damping ratio of \(\gamma=0.07\). The results demonstrate good agreement between the analytical predictions and the FEA.}
\label{fig:Fig1}
\end{figure}

Initially, we consider a straight beam. We apply a compressive axial load that exceeds the critical buckling load, \(\hat{P}_{\text{cr}}\) (see Eq.~\eqref{eq:1}), to the left end as depicted in Fig.~\ref{fig:Fig1}(a),
\begin{equation}\label{eq:1}
\hat{P}_{\text{cr}} = \frac{4\pi^2EI}{L^2},
\end{equation} 
where \(L\), \(E\), and \(I = \frac{bt^3}{12}\) represent the length, Young's modulus, and moment of inertia of the buckled beam, respectively. The compressive axial load buckles the beam to a static deflection position, defined by Eq.~\eqref{eq:2}~\cite{timoshenko2009theory},
\begin{equation}\label{eq:2}
{\hat{W}_o}(\hat{x}) = \frac{h}{2} \left\{1 - \cos \left(2\pi\frac{\hat{x}}{L}\right) \right\}.
\end{equation}

We then fix both ends of the beam, resulting in a curved beam, as shown in Fig.~\ref{fig:Fig1}(b). The initial static deflection of the beam is defined by Eq.\,\eqref{eq:2} and the vertical rise of the midpoint is given by $\hat W_o(x=\frac{L}{2})= h$, where $h$ represents the amplitude of the first buckled mode shape of the clamped-clamped beam,

The beam has two stable states (Fig.~\ref{fig:Fig1}(b)). We can induce a switch between its stable states by applying harmonic excitation at the midpoint, given by $\hat{W}_o(x=\frac{L}{2})= h$, as illustrated in Fig.\,\ref{fig:Fig1}(b). To model the bistable buckled beam analytically, we begin with the nonlinear Euler-Bernoulli beam equation, which is represented by Eq.\,\eqref{eq:3}~\cite{tseng1971nonlinear}, 
\begin{equation} \label{eq:3}
\begin{aligned}
& EI \frac{\partial^4 \hat{W}}{\partial \hat{x}^4} + \hat P_{cr}\frac{\partial^2 \hat{W}}{\partial \hat{x}^2} + \rho A \frac{\partial^2 \hat{W}}{\partial \hat{t}^2} + \hat{C}^d \frac{\partial \hat{W}}{\partial \hat{t}} \\
& - \frac{EA}{2L} \left\{\int_0^L \left[\left(\frac{\partial \hat{W}}{\partial \hat{x}}\right)^2 + 2\frac{\partial \hat{W}}{\partial \hat{x}}\frac{\partial \hat{W}_o}{\partial \hat{x}}\right] d\hat{x}\right\} \\
& \times \left( \frac{\partial^2 \hat{W}}{\partial\hat{x}^2}+\frac{\partial^2\hat{W}_o}{\partial\hat{x}^2}\right) = \hat F \cos{(\hat{\Omega} \hat{t})}.
\end{aligned}
\end{equation}
By approximating the first buckling mode using Eq.\,\eqref{eq:2} and applying Galerkin approximation, we discretize Eq.\,\eqref{eq:3} into Eq.\,\eqref{eq:4}, which represents the symmetric bistable Duffing equation~\cite{SI} (see supplemental materials~\cite{SI} for detailed derivation). 
\begin{equation}\label{eq:4}
\begin{split}
{\ddot u}+ \gamma {\dot u}- u+u^3 
= G\cos{(\Omega \tau)},
\end{split}
\end{equation}
where \(G\), \(\Omega\), \(\gamma\) are the non-dimensionalized forcing amplitude, excitation frequency, and damping ratio, respectively (see supplemental materials~\cite{SI} for derivations). Equation~\eqref{eq:4}  depicts the dimensionless symmetric bistable Duffing equation, characterized by double-well potential with two stable equilibrium points at $u_{-1} = -1$ and $u_{+1} = +1$. These points are separated by an unstable equilibrium, or ``hilltop", at $u_0 = 0$~\cite{PhysRevE.108.L022201}.
Figure~\ref{fig:Fig1}(c) illustrates four distinct behaviors: switching, reverting, vacillating, and intra-well. Our previous study established numerical criteria to distinguish among these behaviors by conducting time-domain simulations on Eq.~\eqref{eq:4}~\cite{PhysRevE.108.L022201}. We fix the initial conditions in all simulations at $(u,\dot u)=(-1,0)$. Here, we verify four distinct behaviors by conducting finite element analysis using Abaqus/Standard. We normalize all the dimensions using the initial vertical rise of the beam  $\hat W_o(x=\frac{L}{2})= h=5.22$ mm. Initially, we model a straight beam with normalized dimensions: length \(L/h = 11.5\), thickness \(t/h = 0.2\), and width \(b/h = 1.92\). This beam undergoes buckling when subjected to a load exceeding the critical buckling load, which is \(\frac{\hat P_{\text{cr}}L^2}{EI} = 39.478\) (see supplemental materials~\cite{SI} for detailed derivation). Subsequently, we perform a modal analysis on the buckled beam in Abaqus/Standard to identify its first buckling mode. We use B21 elements and a hyperelastic material, Dragon Skin 30~\cite{smooth-on}, which is nearly incompressible (Poisson's ratio \(\nu \approx 0.495\)), with an initial Young's modulus of \(E = 0.74 \pm 0.07\) MPa~\cite{ranzani2015bioinspired}, to model the bistable buckled beam.

Then, we perform dynamic implicit analysis to confirm the four types of behavior predicted analytically: switching behavior with dimensionless parameters \((G,\Omega)=(0.100,1.17595)\), reverting behavior with parameters \((G,\Omega)=(0.125,1.209412)\), aperiodic vacillating behavior with parameters \((G,\Omega)=(0.175,1.13291)\), and intra-well behavior with parameters \((G,\Omega)=(0.150,1.30506)\), all under a constant damping ratio of \(\gamma=0.07\). Figure~\ref{fig:Fig1}(c) demonstrates good agreement between analytical and FEA results for these behaviors. 
The static component of Eq.~\eqref{eq:4}, \( F_{\text{static}} = -u + u^3 \), characterizes an 
energy landscape with equal potential wells, requiring a substantial static actuation force of \(0.38\) units to transition the bistable buckled beam from one stable state to another~\cite{PhysRevE.108.L022201}. Figure~\ref{fig:Fig1}(c) demonstrates that dynamic excitation of the bistable system can 
significantly reduce the required actuation force compared with quasi-static actuation~\cite{PhysRevE.108.L022201}. In a previous study, we presented the forcing amplitude-frequency parameter space for a symmetric bistable Duffing system at a constant damping ratio of \(\gamma = 0.07\)~\cite{PhysRevE.108.L022201}. The findings revealed that although low forcing amplitudes can produce four distinct behaviors, the simultaneous existence of switching, reverting, and vacillating behaviors near each other in this parameter space leads to an intertwined chaotic region. The intertwined chaotic region in the parameter space complicates the selection of an appropriate switching frequency and forcing amplitude for transitioning the bistable buckled beam between its stable states.

%Recent studies have demonstrated that applying a magnetic field can programmatically alter the energy landscape of a magnetized bistable buckled beam. Static analysis of such beams under varying magnetic fields shows that it is possible to tune stability characteristics and modify the energy landscape. In particular, the force-displacement curve experiences a linear translation toward one of the stable states, dictated by the direction of the applied magnetic field~\cite{pal2023programmable,abbasi2023snap}. 

%Here, we apply static bias force \(P\textrm{(B)}\) to linearly shift the force-displacement curve along the force axis. Essentially, \(P\textrm{(B)}\) quantifies the magnitude of the shift in the force-displacement curve along the force axis and leads to a change in the system's energy landscape.

%Recent studies have demonstrated that applying a magnetic field can programmatically alter the energy landscape of a magnetized bistable buckled beam. Static analysis of such beams under varying magnetic fields shows that it is possible to tune stability characteristics and modify the energy landscape. 
Recent studies have demonstrated that applying a magnetic field can programmatically alter the energy landscape of a magnetized bistable buckled beam, allowing for the tuning of stability characteristics and modification of the energy landscape through static analysis under varying magnetic fields~\cite{abbasi2023snap,pal2023programmable,zhao2023tuning,zhao2023encoding}. In particular, the force-displacement curve experiences a linear translation toward one of the stable states, dictated by the direction of the applied magnetic field~\cite{pal2023programmable,abbasi2023snap}.

\begin{figure}[t]
\centering
\includegraphics[scale=0.65]{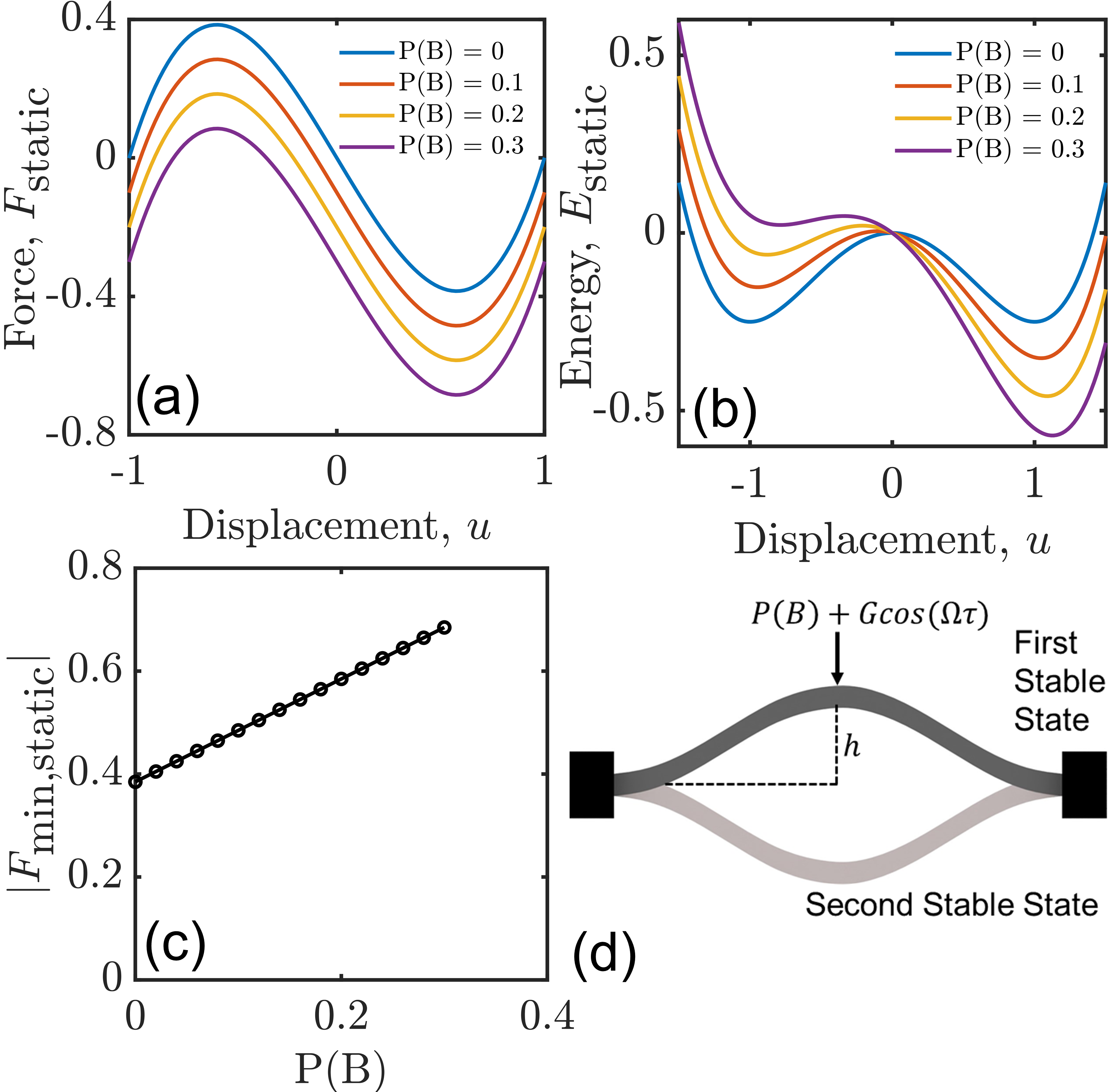}
\caption{\label{fig:Fig2}
Tuning stability characteristics of force-displacement and energy landscapes under varying static bias forces \(P\textrm{(B)}\): (a) Evolution of the force-displacement curve, showing linear shifts along the force axis in response to varying levels of static bias force, \(P\textrm{(B)}\). (b) The transition from symmetric bistable to asymmetric bistable energy landscapes with increasing static bias force. (c) A linear relationship between the absolute value of \(F_{\text{min,static}}\) and the static bias force \(P\textrm{(B)}\). (d) A demonstration of combined static bias force and dynamic excitation to the bistable buckled beam.}
\end{figure}

%Building on this concept, we apply a static bias force \(P\textrm{(B)}\) to linearly shift the force-displacement curve along the force axis. Essentially, \(P\textrm{(B)}\) quantifies the magnitude of the shift in the force-displacement curve along the force axis and leads to a change in the system's energy landscape. By adjusting the static bias force, we can similarly manipulate the stability characteristics and energy landscape of the beam, achieving a controlled shift similar to the effects observed with magnetic fields. 
Building on this concept, we apply a static bias force \(P\textrm{(B)}\) to linearly shift the force-displacement curve along the force axis, altering the system's energy landscape. By adjusting \(P\textrm{(B)}\), we can manipulate the stability characteristics of the beam, achieving a controlled shift similar to the effects observed with magnetic fields.
Incorporating \(P\textrm{(B)}\) as a static bias force changes Eq.~\eqref{eq:4} to Eq.~\eqref{eq:5}, altering the system's response. Equation~\eqref{eq:5} is the dimensionless asymmetric bistable Duffing equation, characterized by an asymmetric double-well potential with two stable equilibrium points.
The impact of the static bias force \(P\textrm{(B)}\) on the symmetric bistable system is understood through the static component of Eq.~\eqref{eq:5}, \(F_{\text{static}} = -u + u^3 - P\textrm{(B)}\). This influence is visually demonstrated in Figs.~\ref{fig:Fig2}(a) and (b), which show how the force-displacement and energy-displacement curves change with the application of \(P\textrm{(B)}\) (see supplemental materials~\cite{SI} for more results). Specifically, the potential well at the stable equilibrium point \( u_{-1} = -1 \) increases, while the potential well at \( u_{+1} = +1 \) decreases.
%This influence is visually demonstrated in Figs.~\ref{fig:Fig2}(a) and (b), which show the linear translation of both the force-displacement and energy-displacement curves with the application of \(P\textrm{(B)}\) (see supplemental materials~\cite{SI} for more results). 
Figure~\ref{fig:Fig2}(c) presents the absolute value of \(F_{\text{min,static}}\), the minimum force in the second stable state, for each force-displacement curve across varying values of \(P\textrm{(B)}\), demonstrating a linear relationship between absolute value of \(F_{\text{min,static}}\) and the static bias force \(P\textrm{(B)}\). Now, we can apply a combined static bias force with dynamic excitation to switch the bistable buckled beam from one stable state to another, as demonstrated in Fig.~\ref{fig:Fig2}(d).
\begin{equation}\label{eq:5}
\begin{split}
{\ddot u}+ \gamma {\dot u}- u+u^3 
= G\cos{(\Omega \tau)}+P\textrm{(B)},
\end{split}
\end{equation}

\begin{figure}[t]
\centering
\includegraphics[width=\linewidth]{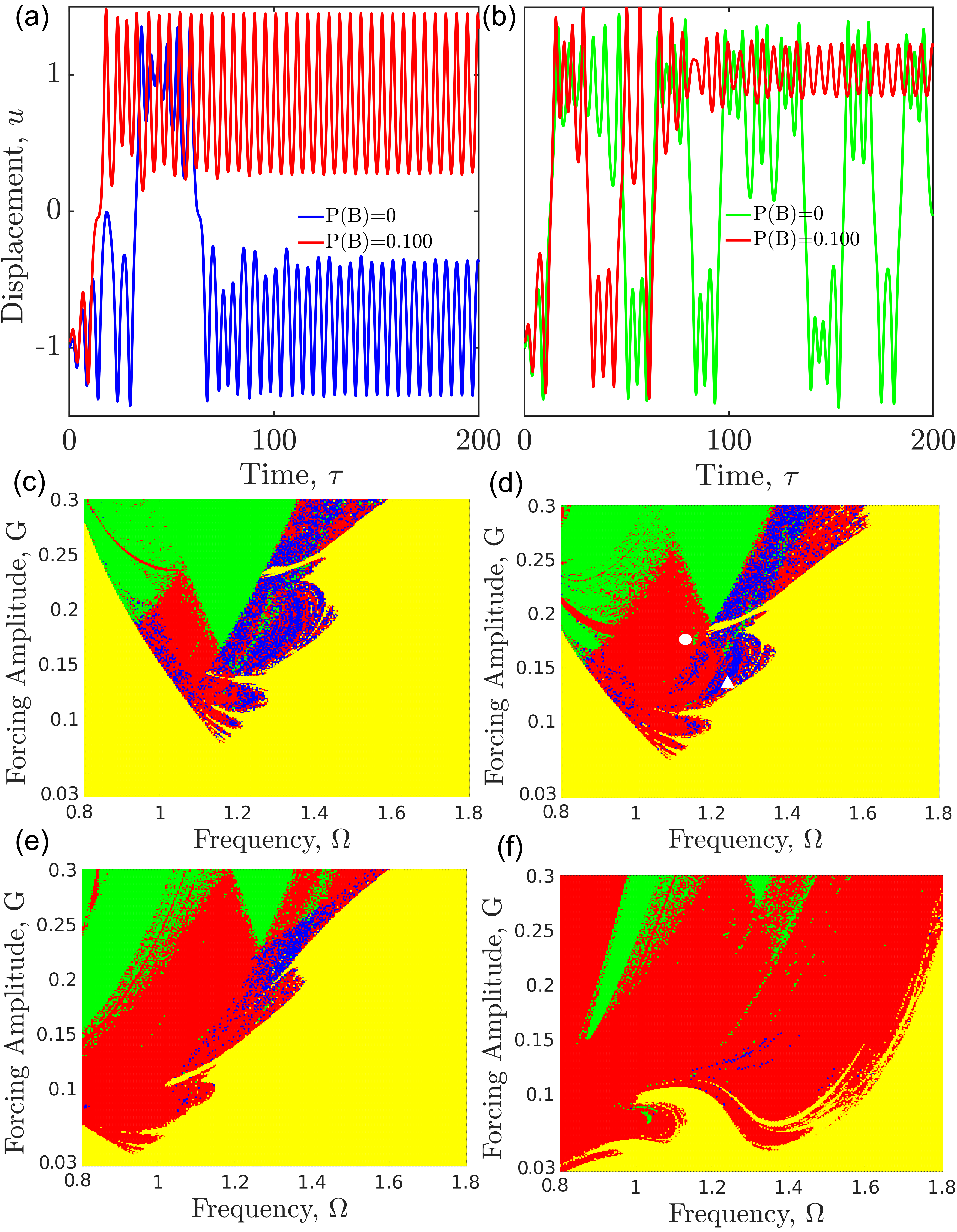}
\caption{Morphing of the bistable buckled beam's behavior due to combined dynamic excitation and static bias force: (a) The reverting behavior depicted in Fig.~\ref{fig:Fig1}(c) transitions to switching behavior when subjected to both dynamic excitation and a static bias force. (b) Similarly, the vacillating behavior shown in Fig.~\ref{fig:Fig1}(c) shifts to switching behavior under the same combined forces. Panels (c), (d), (e), and (f) display the forcing amplitude-frequency parameter space for the bistable buckled beam with a damping ratio of \(\gamma = 0.07\) and static bias forces of \(P\textrm{(B)} = 0.04, 0.100, 0.200, 0.300\) respectively. The light gray \(\bigcirc\) and \(\triangle\) in Fig.~\ref{fig:Fig3}(d) indicate the points where the reverting behavior with parameters \((G, \Omega) = (0.125, 1.209412)\) and the aperiodic vacillating behavior with parameters \((G, \Omega) = (0.175, 1.13291)\) under \(P(\textrm{B}) = 0.100\) transition into switching behavior.}
\label{fig:Fig3}
\end{figure}

We next conduct time-domain simulation using the fourth-order Runge-Kutta scheme to solve Eq.~\eqref{eq:5} for $P\textrm{(B)}=0.100$ with parameters of $(G,\Omega)=(0.125,1.209412)$ of reverting behavior of Fig.~\ref{fig:Fig1}(c) and a damping ratio of $\gamma=0.07$. Fig.~\ref{fig:Fig3}(a) presents the simulation results, demonstrating that under the influence of a static bias force, the reverting behavior depicted in Fig.~\ref{fig:Fig1}(c) transitions to switching behavior. Similarly, for vacillating behavior with parameters $(G,\Omega)=(0.175,1.13291)$ and a damping ratio of $\gamma=0.07$, the static bias force of $P\textrm{(B)}=0.100$ changes the vacillating behavior of Fig.~\ref{fig:Fig1}(c) into switching behavior, as shown in Fig.~\ref{fig:Fig3}(b).
Figures~\ref{fig:Fig3}(c)-(f) display the forcing amplitude-frequency parameter space within the ranges \(0.80 \leq \Omega \leq 1.8\) and \(0.03 \leq G \leq 0.30\), with a damping ratio of \(\gamma=0.07\) and static bias forces \(P\textrm{(B)}=0.04, 0.100, 0.200, 0.300\) respectively. For each combination of \((G, \Omega)\), we conduct time-domain simulations of Eq.~\eqref{eq:5} across these parameter ranges, incorporating the aforementioned damping ratios and static bias forces. Figures~\ref{fig:Fig3}(c)-(f) present the results for \(P\textrm{(B)} = 0.04, 0.100, 0.200, 0.300\), plotted on a \(256\times256\) grid (for additional results with other \(P\textrm{(B)}\) values, see supplemental materials~\cite{SI}). We classify each simulation's numerical steady state into four behaviors: switching, reverting, vacillating, or intra-well, represented by red, blue, green, and yellow data points in Figs.~\ref{fig:Fig3}(c)-(f), respectively. This categorization follows the methodology established in our previous study~\cite{PhysRevE.108.L022201}. One notable observation from Figs.~\ref{fig:Fig3}(c)-(f) is that as the static bias force increases, the switching behavior becomes more prominent across all parameter spaces compared to other behaviors. Furthermore, as \(P\textrm{(B)}\) increases, the minimum dynamic forcing amplitude required to switch between stable states in both directions decreases. This means that a higher static bias not only makes the potential well asymmetric but also reduces the dynamic forcing amplitude requirement of \(G_\textrm{min}\). Here, \(G_\textrm{min}\) denotes the minimum forcing amplitude required for switching behavior. \(P\textrm{(B)}\) and \(G_\textrm{min}\) have a linear relationship, with \(G_\textrm{min}\) decreasing linearly as \(P\textrm{(B)}\) increases (see supplemental materials~\cite{SI} for the relationship between \(P\textrm{(B)}\) and \(G_\textrm{min}\)).

\begin{figure}[t!]
\centering
\includegraphics[width=\linewidth]{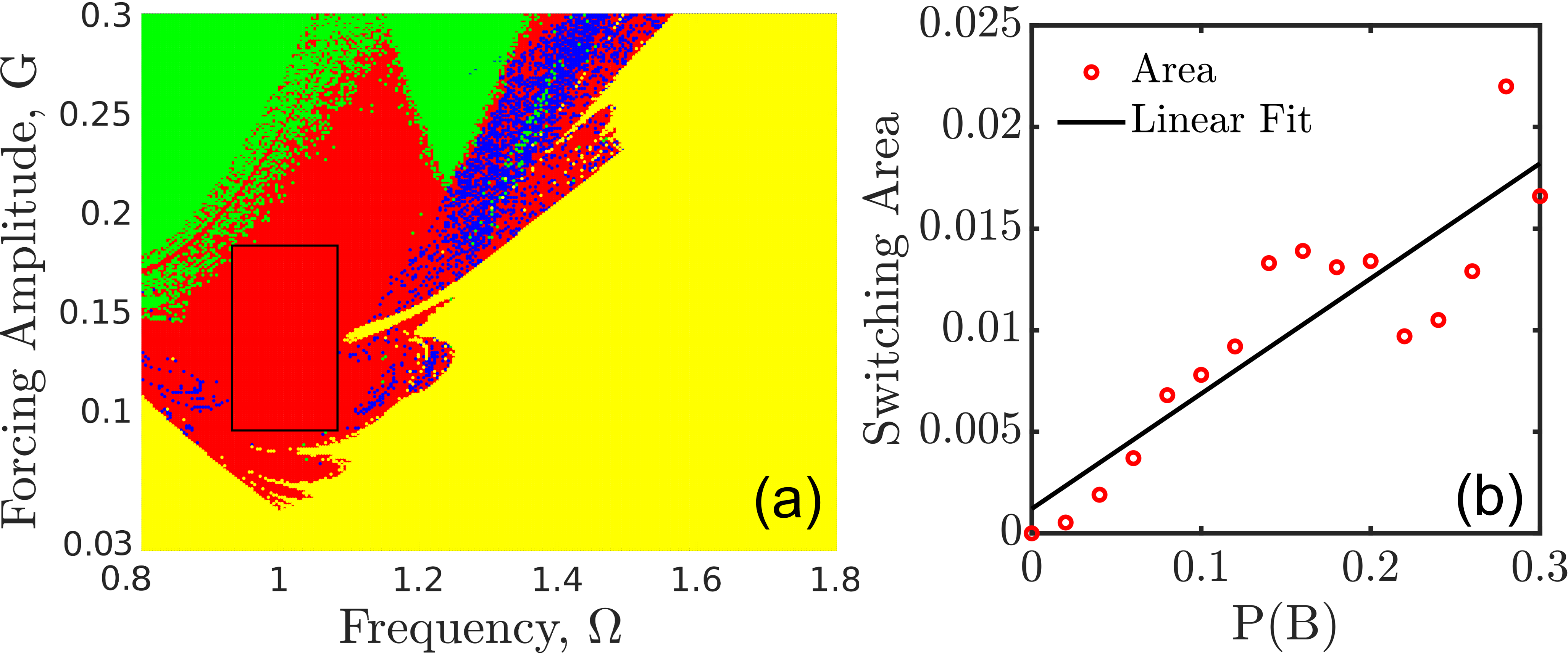}
\caption{Distinct switching areas: (a) shows a parameter space for \(P\textrm{(B)}=0.160\) and \(\gamma=0.07\), outlined by a rectangle, where every combination of \(G\) and \(\Omega\) results in switching behavior. (b) illustrates the expansion of the switching behavior area as the static bias force \(P\textrm{(B)}\) increases across different parameter spaces.}
\label{fig:Fig4}
\end{figure}

Next, we quantify the switching behavior area within the parameter spaces of Figs.~\ref{fig:Fig3}(c)-(f)  where the parameter set \((G, \Omega)\) always results in switching behavior. Figure~\ref{fig:Fig4}(a) illustrates a parameter space with static bias force \(P\textrm{(B)}=0.160\) where a rectangle indicates the area where no behaviors other than switching are present. We quantify this area from a \(256 \times 256\) grid, where each set of (\(G, \Omega\)) results in switching behaviors. We numerically detect the largest possible rectangle in the parameter space, excluding other behaviors. After measuring the width and height of the rectangle across the frequency and forcing amplitude ranges, we calculate the distinct dimensionless switching area, \(\Delta \Omega \times \Delta G\), where no other behaviors are present. Figure~\ref{fig:Fig4}(b) shows that as we increase the static bias force, the dimensionless rectangular area in the parameter spaces increases (see supplemental materials~\cite{SI} for additional results).

%\NH{Furthermore, we observe a linear relationship between \( G_\textrm{min} \) and the applied bias force \( P\textrm{(B)} \). Here, \( G_\textrm{min} \) represents the minimum dynamic forcing amplitude required to switch between stable states for different values of \( P\textrm{(B)} \), indicating that as \( P\textrm{(B)} \) increases, the required forcing amplitude decreases (see supplemental materials~\cite{SI} for additional results).}

In conclusion, our investigation shows the dynamic morphing capabilities of bistable buckled beams under the influence of static bias forces and dynamic excitation. We demonstrate that applying a static bias force expands the parameter space conducive to switching behavior, thus facilitating more efficient transitions between stable states without the onset of chaos. This enhancement of the switching behavior region underscores the potential of static bias force as a tool for optimizing the morphing efficiency of bistable systems. %Our findings advance the theoretical understanding of bistable beam dynamics and pave the way for innovative applications in adaptive multi-stable mechanical metamaterials.

\section*{Supplementary Material}
In the supplemental materials, we have included a detailed derivation of the bistable Duffing equation from the nonlinear vibration equation of the buckled beam. Furthermore, we have added additional results of Fig.~\ref{fig:Fig2}, Fig.~\ref{fig:Fig3}, and Fig.~\ref{fig:Fig4}.

%\underline{\textit{Acknowledgements:}} 
We acknowledge the support from the National Institutes of Health (NIH): Project No.\,R01EB032959. Start-up funds from the Department of Mechanical Engineering at the Univ.\,of\,Utah also supported this work. The support and resources from the Center for High-Performance Computing at Univ.\,of\,Utah are gratefully acknowledged. 

\nocite{*}
\bibliography{aipsamp}% Produces the bibliography via BibTeX.

%merlin.mbs aipnum4-1.bst 2010-07-25 4.21a (PWD, AO, DPC) hacked
%Control: key (0)
%Control: author (8) initials jnrlst
%Control: editor formatted (1) identically to author
%Control: production of article title (0) allowed
%Control: page (1) range
%Control: year (1) truncated
%Control: production of eprint (0) enabled
\begin{thebibliography}{51}%
\makeatletter
\providecommand \@ifxundefined [1]{%
 \@ifx{#1\undefined}
}%
\providecommand \@ifnum [1]{%
 \ifnum #1\expandafter \@firstoftwo
 \else \expandafter \@secondoftwo
 \fi
}%
\providecommand \@ifx [1]{%
 \ifx #1\expandafter \@firstoftwo
 \else \expandafter \@secondoftwo
 \fi
}%
\providecommand \natexlab [1]{#1}%
\providecommand \enquote  [1]{``#1''}%
\providecommand \bibnamefont  [1]{#1}%
\providecommand \bibfnamefont [1]{#1}%
\providecommand \citenamefont [1]{#1}%
\providecommand \href@noop [0]{\@secondoftwo}%
\providecommand \href [0]{\begingroup \@sanitize@url \@href}%
\providecommand \@href[1]{\@@startlink{#1}\@@href}%
\providecommand \@@href[1]{\endgroup#1\@@endlink}%
\providecommand \@sanitize@url [0]{\catcode `\\12\catcode `\$12\catcode `\&12\catcode `\#12\catcode `\^12\catcode `\_12\catcode `\%12\relax}%
\providecommand \@@startlink[1]{}%
\providecommand \@@endlink[0]{}%
\providecommand \url  [0]{\begingroup\@sanitize@url \@url }%
\providecommand \@url [1]{\endgroup\@href {#1}{\urlprefix }}%
\providecommand \urlprefix  [0]{URL }%
\providecommand \Eprint [0]{\href }%
\providecommand \doibase [0]{http://dx.doi.org/}%
\providecommand \selectlanguage [0]{\@gobble}%
\providecommand \bibinfo  [0]{\@secondoftwo}%
\providecommand \bibfield  [0]{\@secondoftwo}%
\providecommand \translation [1]{[#1]}%
\providecommand \BibitemOpen [0]{}%
\providecommand \bibitemStop [0]{}%
\providecommand \bibitemNoStop [0]{.\EOS\space}%
\providecommand \EOS [0]{\spacefactor3000\relax}%
\providecommand \BibitemShut  [1]{\csname bibitem#1\endcsname}%
\let\auto@bib@innerbib\@empty
%</preamble>
\bibitem [{\citenamefont {Hasan}\ \emph {et~al.}(2023{\natexlab{a}})\citenamefont {Hasan}, \citenamefont {Greenwood}, \citenamefont {Kong},\ and\ \citenamefont {Wang}}]{hasan2023optimization}%
  \BibitemOpen
  \bibfield  {author} {\bibinfo {author} {\bibfnamefont {M.~N.}\ \bibnamefont {Hasan}}, \bibinfo {author} {\bibfnamefont {T.}~\bibnamefont {Greenwood}}, \bibinfo {author} {\bibfnamefont {Y.~L.}\ \bibnamefont {Kong}}, \ and\ \bibinfo {author} {\bibfnamefont {P.}~\bibnamefont {Wang}},\ }\bibfield  {title} {\enquote {\bibinfo {title} {Optimization of the small-amplitude dynamic triggering mechanism of bi-stable metamaterials},}\ }\href@noop {} {\bibfield  {journal} {\bibinfo  {journal} {Bull. Am. Phys. Soc.}\ }\textbf {\bibinfo {volume} {68}} (\bibinfo {year} {2023}{\natexlab{a}})}\BibitemShut {NoStop}%
\bibitem [{\citenamefont {Shan}\ \emph {et~al.}(2015)\citenamefont {Shan}, \citenamefont {Kang}, \citenamefont {Raney}, \citenamefont {Wang}, \citenamefont {Fang}, \citenamefont {Candido}, \citenamefont {Lewis},\ and\ \citenamefont {Bertoldi}}]{shan2015multistable}%
  \BibitemOpen
  \bibfield  {author} {\bibinfo {author} {\bibfnamefont {S.}~\bibnamefont {Shan}}, \bibinfo {author} {\bibfnamefont {S.~H.}\ \bibnamefont {Kang}}, \bibinfo {author} {\bibfnamefont {J.~R.}\ \bibnamefont {Raney}}, \bibinfo {author} {\bibfnamefont {P.}~\bibnamefont {Wang}}, \bibinfo {author} {\bibfnamefont {L.}~\bibnamefont {Fang}}, \bibinfo {author} {\bibfnamefont {F.}~\bibnamefont {Candido}}, \bibinfo {author} {\bibfnamefont {J.~A.}\ \bibnamefont {Lewis}}, \ and\ \bibinfo {author} {\bibfnamefont {K.}~\bibnamefont {Bertoldi}},\ }\bibfield  {title} {\enquote {\bibinfo {title} {Multistable architected materials for trapping elastic strain energy},}\ }\href@noop {} {\bibfield  {journal} {\bibinfo  {journal} {Adv. Mater.}\ }\textbf {\bibinfo {volume} {27}},\ \bibinfo {pages} {4296--4301} (\bibinfo {year} {2015})}\BibitemShut {NoStop}%
\bibitem [{\citenamefont {Mei}\ \emph {et~al.}(2021)\citenamefont {Mei}, \citenamefont {Meng}, \citenamefont {Zhao},\ and\ \citenamefont {Chen}}]{mei2021mechanical}%
  \BibitemOpen
  \bibfield  {author} {\bibinfo {author} {\bibfnamefont {T.}~\bibnamefont {Mei}}, \bibinfo {author} {\bibfnamefont {Z.}~\bibnamefont {Meng}}, \bibinfo {author} {\bibfnamefont {K.}~\bibnamefont {Zhao}}, \ and\ \bibinfo {author} {\bibfnamefont {C.~Q.}\ \bibnamefont {Chen}},\ }\bibfield  {title} {\enquote {\bibinfo {title} {A mechanical metamaterial with reprogrammable logical functions},}\ }\href@noop {} {\bibfield  {journal} {\bibinfo  {journal} {Nat. Commun.}\ }\textbf {\bibinfo {volume} {12}},\ \bibinfo {pages} {7234} (\bibinfo {year} {2021})}\BibitemShut {NoStop}%
\bibitem [{\citenamefont {Gupta}, \citenamefont {Adhikari},\ and\ \citenamefont {Bhattacharya}(2024)}]{gupta2024evidence}%
  \BibitemOpen
  \bibfield  {author} {\bibinfo {author} {\bibfnamefont {V.}~\bibnamefont {Gupta}}, \bibinfo {author} {\bibfnamefont {S.}~\bibnamefont {Adhikari}}, \ and\ \bibinfo {author} {\bibfnamefont {B.}~\bibnamefont {Bhattacharya}},\ }\bibfield  {title} {\enquote {\bibinfo {title} {Evidence of nonlinearity tailoring in static and dynamic responses of honeycomb and auxetic hourglass lattice metastructures},}\ }\href@noop {} {\bibfield  {journal} {\bibinfo  {journal} {Mech. Res. Commun.}\ }\textbf {\bibinfo {volume} {137}},\ \bibinfo {pages} {104261} (\bibinfo {year} {2024})}\BibitemShut {NoStop}%
\bibitem [{\citenamefont {de~Castro}, \citenamefont {Orlando},\ and\ \citenamefont {Gon{\c{c}}alves}(2024)}]{de2024nonlinear}%
  \BibitemOpen
  \bibfield  {author} {\bibinfo {author} {\bibfnamefont {C.}~\bibnamefont {de~Castro}}, \bibinfo {author} {\bibfnamefont {D.}~\bibnamefont {Orlando}}, \ and\ \bibinfo {author} {\bibfnamefont {P.}~\bibnamefont {Gon{\c{c}}alves}},\ }\bibfield  {title} {\enquote {\bibinfo {title} {Nonlinear dynamics of elastically connected multistable structures},}\ }in\ \href@noop {} {\emph {\bibinfo {booktitle} {J. Phys. Conf. Ser.}}},\ Vol.\ \bibinfo {volume} {2647}\ (\bibinfo {organization} {IOP Publishing},\ \bibinfo {year} {2024})\ p.\ \bibinfo {pages} {092001}\BibitemShut {NoStop}%
\bibitem [{\citenamefont {And{\`o}}\ \emph {et~al.}(2014)\citenamefont {And{\`o}}, \citenamefont {Baglio}, \citenamefont {Bulsara},\ and\ \citenamefont {Marletta}}]{ando2014bistable}%
  \BibitemOpen
  \bibfield  {author} {\bibinfo {author} {\bibfnamefont {B.}~\bibnamefont {And{\`o}}}, \bibinfo {author} {\bibfnamefont {S.}~\bibnamefont {Baglio}}, \bibinfo {author} {\bibfnamefont {A.}~\bibnamefont {Bulsara}}, \ and\ \bibinfo {author} {\bibfnamefont {V.}~\bibnamefont {Marletta}},\ }\bibfield  {title} {\enquote {\bibinfo {title} {A bistable buckled beam based approach for vibrational energy harvesting},}\ }\href@noop {} {\bibfield  {journal} {\bibinfo  {journal} {Sens. Actuators A: Phys.}\ }\textbf {\bibinfo {volume} {211}},\ \bibinfo {pages} {153--161} (\bibinfo {year} {2014})}\BibitemShut {NoStop}%
\bibitem [{\citenamefont {Harne}\ and\ \citenamefont {Wang}(2013)}]{harne2013review}%
  \BibitemOpen
  \bibfield  {author} {\bibinfo {author} {\bibfnamefont {R.~L.}\ \bibnamefont {Harne}}\ and\ \bibinfo {author} {\bibfnamefont {K.}~\bibnamefont {Wang}},\ }\bibfield  {title} {\enquote {\bibinfo {title} {A review of the recent research on vibration energy harvesting via bistable systems},}\ }\href@noop {} {\bibfield  {journal} {\bibinfo  {journal} {Smart Mater. Struct.}\ }\textbf {\bibinfo {volume} {22}},\ \bibinfo {pages} {023001} (\bibinfo {year} {2013})}\BibitemShut {NoStop}%
\bibitem [{\citenamefont {Ibrahim}, \citenamefont {Towfighian},\ and\ \citenamefont {Younis}(2017)}]{ibrahim2017dynamics}%
  \BibitemOpen
  \bibfield  {author} {\bibinfo {author} {\bibfnamefont {A.}~\bibnamefont {Ibrahim}}, \bibinfo {author} {\bibfnamefont {S.}~\bibnamefont {Towfighian}}, \ and\ \bibinfo {author} {\bibfnamefont {M.~I.}\ \bibnamefont {Younis}},\ }\bibfield  {title} {\enquote {\bibinfo {title} {Dynamics of transition regime in bistable vibration energy harvesters},}\ }\href@noop {} {\bibfield  {journal} {\bibinfo  {journal} {J. Vib. Acoust.}\ }\textbf {\bibinfo {volume} {139}},\ \bibinfo {pages} {051008} (\bibinfo {year} {2017})}\BibitemShut {NoStop}%
\bibitem [{\citenamefont {Harne}\ and\ \citenamefont {Wang}(2017)}]{harne2017harnessing}%
  \BibitemOpen
  \bibfield  {author} {\bibinfo {author} {\bibfnamefont {R.~L.}\ \bibnamefont {Harne}}\ and\ \bibinfo {author} {\bibfnamefont {K.-W.}\ \bibnamefont {Wang}},\ }\href@noop {} {\emph {\bibinfo {title} {Harnessing bistable structural dynamics: for vibration control, energy harvesting and sensing}}}\ (\bibinfo  {publisher} {John Wiley \& Sons},\ \bibinfo {year} {2017})\BibitemShut {NoStop}%
\bibitem [{\citenamefont {Yang}\ and\ \citenamefont {Cao}(2020)}]{yang2020dynamics}%
  \BibitemOpen
  \bibfield  {author} {\bibinfo {author} {\bibfnamefont {T.}~\bibnamefont {Yang}}\ and\ \bibinfo {author} {\bibfnamefont {Q.}~\bibnamefont {Cao}},\ }\bibfield  {title} {\enquote {\bibinfo {title} {Dynamics and high-efficiency of a novel multi-stable energy harvesting system},}\ }\href@noop {} {\bibfield  {journal} {\bibinfo  {journal} {Chaos Solitons Fractals}\ }\textbf {\bibinfo {volume} {131}},\ \bibinfo {pages} {109516} (\bibinfo {year} {2020})}\BibitemShut {NoStop}%
\bibitem [{\citenamefont {Chen}, \citenamefont {Pauly},\ and\ \citenamefont {Reis}(2021)}]{chen2021reprogrammable}%
  \BibitemOpen
  \bibfield  {author} {\bibinfo {author} {\bibfnamefont {T.}~\bibnamefont {Chen}}, \bibinfo {author} {\bibfnamefont {M.}~\bibnamefont {Pauly}}, \ and\ \bibinfo {author} {\bibfnamefont {P.~M.}\ \bibnamefont {Reis}},\ }\bibfield  {title} {\enquote {\bibinfo {title} {A reprogrammable mechanical metamaterial with stable memory},}\ }\href@noop {} {\bibfield  {journal} {\bibinfo  {journal} {Nature}\ }\textbf {\bibinfo {volume} {589}},\ \bibinfo {pages} {386--390} (\bibinfo {year} {2021})}\BibitemShut {NoStop}%
\bibitem [{\citenamefont {Meng}\ \emph {et~al.}(2022)\citenamefont {Meng}, \citenamefont {Liu}, \citenamefont {Yan}, \citenamefont {Genin},\ and\ \citenamefont {Chen}}]{meng2022deployable}%
  \BibitemOpen
  \bibfield  {author} {\bibinfo {author} {\bibfnamefont {Z.}~\bibnamefont {Meng}}, \bibinfo {author} {\bibfnamefont {M.}~\bibnamefont {Liu}}, \bibinfo {author} {\bibfnamefont {H.}~\bibnamefont {Yan}}, \bibinfo {author} {\bibfnamefont {G.~M.}\ \bibnamefont {Genin}}, \ and\ \bibinfo {author} {\bibfnamefont {C.~Q.}\ \bibnamefont {Chen}},\ }\bibfield  {title} {\enquote {\bibinfo {title} {Deployable mechanical metamaterials with multistep programmable transformation},}\ }\href@noop {} {\bibfield  {journal} {\bibinfo  {journal} {Sci. Adv.}\ }\textbf {\bibinfo {volume} {8}},\ \bibinfo {pages} {eabn5460} (\bibinfo {year} {2022})}\BibitemShut {NoStop}%
\bibitem [{\citenamefont {Restrepo}, \citenamefont {Mankame},\ and\ \citenamefont {Zavattieri}(2015)}]{restrepo2015phase}%
  \BibitemOpen
  \bibfield  {author} {\bibinfo {author} {\bibfnamefont {D.}~\bibnamefont {Restrepo}}, \bibinfo {author} {\bibfnamefont {N.~D.}\ \bibnamefont {Mankame}}, \ and\ \bibinfo {author} {\bibfnamefont {P.~D.}\ \bibnamefont {Zavattieri}},\ }\bibfield  {title} {\enquote {\bibinfo {title} {Phase transforming cellular materials},}\ }\href@noop {} {\bibfield  {journal} {\bibinfo  {journal} {Extreme Mech. Lett.}\ }\textbf {\bibinfo {volume} {4}},\ \bibinfo {pages} {52--60} (\bibinfo {year} {2015})}\BibitemShut {NoStop}%
\bibitem [{\citenamefont {Paul}\ \emph {et~al.}(2024)\citenamefont {Paul}, \citenamefont {Overvelde}, \citenamefont {Hochhalter},\ and\ \citenamefont {Wang}}]{paul2024effects}%
  \BibitemOpen
  \bibfield  {author} {\bibinfo {author} {\bibfnamefont {S.}~\bibnamefont {Paul}}, \bibinfo {author} {\bibfnamefont {J.~T.}\ \bibnamefont {Overvelde}}, \bibinfo {author} {\bibfnamefont {J.}~\bibnamefont {Hochhalter}}, \ and\ \bibinfo {author} {\bibfnamefont {P.}~\bibnamefont {Wang}},\ }\bibfield  {title} {\enquote {\bibinfo {title} {Effects of void geometry on two-dimensional monolithic porous phononic crystals},}\ }\href@noop {} {\bibfield  {journal} {\bibinfo  {journal} {Appl. Phys. Lett.}\ }\textbf {\bibinfo {volume} {124}} (\bibinfo {year} {2024})}\BibitemShut {NoStop}%
\bibitem [{\citenamefont {Cao}\ \emph {et~al.}(2021)\citenamefont {Cao}, \citenamefont {Derakhshani}, \citenamefont {Fang}, \citenamefont {Huang},\ and\ \citenamefont {Cao}}]{cao2021bistable}%
  \BibitemOpen
  \bibfield  {author} {\bibinfo {author} {\bibfnamefont {Y.}~\bibnamefont {Cao}}, \bibinfo {author} {\bibfnamefont {M.}~\bibnamefont {Derakhshani}}, \bibinfo {author} {\bibfnamefont {Y.}~\bibnamefont {Fang}}, \bibinfo {author} {\bibfnamefont {G.}~\bibnamefont {Huang}}, \ and\ \bibinfo {author} {\bibfnamefont {C.}~\bibnamefont {Cao}},\ }\bibfield  {title} {\enquote {\bibinfo {title} {Bistable structures for advanced functional systems},}\ }\href@noop {} {\bibfield  {journal} {\bibinfo  {journal} {Adv. Funct. Mater.}\ }\textbf {\bibinfo {volume} {31}},\ \bibinfo {pages} {2106231} (\bibinfo {year} {2021})}\BibitemShut {NoStop}%
\bibitem [{\citenamefont {Qiu}, \citenamefont {Lang},\ and\ \citenamefont {Slocum}(2004)}]{qiu2004curved}%
  \BibitemOpen
  \bibfield  {author} {\bibinfo {author} {\bibfnamefont {J.}~\bibnamefont {Qiu}}, \bibinfo {author} {\bibfnamefont {J.~H.}\ \bibnamefont {Lang}}, \ and\ \bibinfo {author} {\bibfnamefont {A.~H.}\ \bibnamefont {Slocum}},\ }\bibfield  {title} {\enquote {\bibinfo {title} {A curved-beam bistable mechanism},}\ }\href@noop {} {\bibfield  {journal} {\bibinfo  {journal} {J. Microelectromech. Syst.}\ }\textbf {\bibinfo {volume} {13}},\ \bibinfo {pages} {137--146} (\bibinfo {year} {2004})}\BibitemShut {NoStop}%
\bibitem [{\citenamefont {Hasan}, \citenamefont {Alsaleem},\ and\ \citenamefont {Ouakad}(2018)}]{hasan2018novel}%
  \BibitemOpen
  \bibfield  {author} {\bibinfo {author} {\bibfnamefont {M.~H.}\ \bibnamefont {Hasan}}, \bibinfo {author} {\bibfnamefont {F.~M.}\ \bibnamefont {Alsaleem}}, \ and\ \bibinfo {author} {\bibfnamefont {H.~M.}\ \bibnamefont {Ouakad}},\ }\bibfield  {title} {\enquote {\bibinfo {title} {Novel threshold pressure sensors based on nonlinear dynamics of mems resonators},}\ }\href@noop {} {\bibfield  {journal} {\bibinfo  {journal} {J. Micromech. Microeng.}\ }\textbf {\bibinfo {volume} {28}},\ \bibinfo {pages} {065007} (\bibinfo {year} {2018})}\BibitemShut {NoStop}%
\bibitem [{\citenamefont {Pontecorvo}\ \emph {et~al.}(2013)\citenamefont {Pontecorvo}, \citenamefont {Barbarino}, \citenamefont {Murray},\ and\ \citenamefont {Gandhi}}]{pontecorvo2013bistable}%
  \BibitemOpen
  \bibfield  {author} {\bibinfo {author} {\bibfnamefont {M.~E.}\ \bibnamefont {Pontecorvo}}, \bibinfo {author} {\bibfnamefont {S.}~\bibnamefont {Barbarino}}, \bibinfo {author} {\bibfnamefont {G.~J.}\ \bibnamefont {Murray}}, \ and\ \bibinfo {author} {\bibfnamefont {F.~S.}\ \bibnamefont {Gandhi}},\ }\bibfield  {title} {\enquote {\bibinfo {title} {Bistable arches for morphing applications},}\ }\href@noop {} {\bibfield  {journal} {\bibinfo  {journal} {J. Intell. Mater. Syst. Struct.}\ }\textbf {\bibinfo {volume} {24}},\ \bibinfo {pages} {274--286} (\bibinfo {year} {2013})}\BibitemShut {NoStop}%
\bibitem [{\citenamefont {Chi}\ \emph {et~al.}(2022)\citenamefont {Chi}, \citenamefont {Li}, \citenamefont {Zhao}, \citenamefont {Hong}, \citenamefont {Tang},\ and\ \citenamefont {Yin}}]{chi2022bistable}%
  \BibitemOpen
  \bibfield  {author} {\bibinfo {author} {\bibfnamefont {Y.}~\bibnamefont {Chi}}, \bibinfo {author} {\bibfnamefont {Y.}~\bibnamefont {Li}}, \bibinfo {author} {\bibfnamefont {Y.}~\bibnamefont {Zhao}}, \bibinfo {author} {\bibfnamefont {Y.}~\bibnamefont {Hong}}, \bibinfo {author} {\bibfnamefont {Y.}~\bibnamefont {Tang}}, \ and\ \bibinfo {author} {\bibfnamefont {J.}~\bibnamefont {Yin}},\ }\bibfield  {title} {\enquote {\bibinfo {title} {Bistable and multistable actuators for soft robots: Structures, materials, and functionalities},}\ }\href@noop {} {\bibfield  {journal} {\bibinfo  {journal} {Adv. Mater.}\ }\textbf {\bibinfo {volume} {34}},\ \bibinfo {pages} {2110384} (\bibinfo {year} {2022})}\BibitemShut {NoStop}%
\bibitem [{\citenamefont {Librandi}, \citenamefont {Tubaldi},\ and\ \citenamefont {Bertoldi}(2021)}]{librandi2021programming}%
  \BibitemOpen
  \bibfield  {author} {\bibinfo {author} {\bibfnamefont {G.}~\bibnamefont {Librandi}}, \bibinfo {author} {\bibfnamefont {E.}~\bibnamefont {Tubaldi}}, \ and\ \bibinfo {author} {\bibfnamefont {K.}~\bibnamefont {Bertoldi}},\ }\bibfield  {title} {\enquote {\bibinfo {title} {Programming nonreciprocity and reversibility in multistable mechanical metamaterials},}\ }\href@noop {} {\bibfield  {journal} {\bibinfo  {journal} {Nat. Commun.}\ }\textbf {\bibinfo {volume} {12}},\ \bibinfo {pages} {3454} (\bibinfo {year} {2021})}\BibitemShut {NoStop}%
\bibitem [{\citenamefont {Librandi}, \citenamefont {Tubaldi},\ and\ \citenamefont {Bertoldi}(2020)}]{librandi2020snapping}%
  \BibitemOpen
  \bibfield  {author} {\bibinfo {author} {\bibfnamefont {G.}~\bibnamefont {Librandi}}, \bibinfo {author} {\bibfnamefont {E.}~\bibnamefont {Tubaldi}}, \ and\ \bibinfo {author} {\bibfnamefont {K.}~\bibnamefont {Bertoldi}},\ }\bibfield  {title} {\enquote {\bibinfo {title} {Snapping of hinged arches under displacement control: Strength loss and nonreciprocity},}\ }\href@noop {} {\bibfield  {journal} {\bibinfo  {journal} {Phys. Rev. E}\ }\textbf {\bibinfo {volume} {101}},\ \bibinfo {pages} {053004} (\bibinfo {year} {2020})}\BibitemShut {NoStop}%
\bibitem [{\citenamefont {Zhang}\ \emph {et~al.}(2017)\citenamefont {Zhang}, \citenamefont {Zhang}, \citenamefont {Hao}, \citenamefont {Nie},\ and\ \citenamefont {Qiu}}]{zhang2017exploiting}%
  \BibitemOpen
  \bibfield  {author} {\bibinfo {author} {\bibfnamefont {J.}~\bibnamefont {Zhang}}, \bibinfo {author} {\bibfnamefont {C.}~\bibnamefont {Zhang}}, \bibinfo {author} {\bibfnamefont {L.}~\bibnamefont {Hao}}, \bibinfo {author} {\bibfnamefont {R.}~\bibnamefont {Nie}}, \ and\ \bibinfo {author} {\bibfnamefont {J.}~\bibnamefont {Qiu}},\ }\bibfield  {title} {\enquote {\bibinfo {title} {Exploiting the instability of smart structure for reconfiguration},}\ }\href@noop {} {\bibfield  {journal} {\bibinfo  {journal} {Appl. Phys. Lett.}\ }\textbf {\bibinfo {volume} {111}} (\bibinfo {year} {2017})}\BibitemShut {NoStop}%
\bibitem [{\citenamefont {Wang}\ and\ \citenamefont {Frazier}(2023)}]{wang2023phase}%
  \BibitemOpen
  \bibfield  {author} {\bibinfo {author} {\bibfnamefont {C.}~\bibnamefont {Wang}}\ and\ \bibinfo {author} {\bibfnamefont {M.~J.}\ \bibnamefont {Frazier}},\ }\bibfield  {title} {\enquote {\bibinfo {title} {Phase transitions in hierarchical, multi-stable metamaterials},}\ }\href@noop {} {\bibfield  {journal} {\bibinfo  {journal} {Extreme Mech. Lett.}\ }\textbf {\bibinfo {volume} {64}},\ \bibinfo {pages} {102068} (\bibinfo {year} {2023})}\BibitemShut {NoStop}%
\bibitem [{\citenamefont {Li}\ \emph {et~al.}(2021)\citenamefont {Li}, \citenamefont {Avis}, \citenamefont {Chen}, \citenamefont {Wu}, \citenamefont {Zhang}, \citenamefont {Kusumaatmaja},\ and\ \citenamefont {Wang}}]{li2021reconfiguration}%
  \BibitemOpen
  \bibfield  {author} {\bibinfo {author} {\bibfnamefont {Y.}~\bibnamefont {Li}}, \bibinfo {author} {\bibfnamefont {S.~J.}\ \bibnamefont {Avis}}, \bibinfo {author} {\bibfnamefont {J.}~\bibnamefont {Chen}}, \bibinfo {author} {\bibfnamefont {G.}~\bibnamefont {Wu}}, \bibinfo {author} {\bibfnamefont {T.}~\bibnamefont {Zhang}}, \bibinfo {author} {\bibfnamefont {H.}~\bibnamefont {Kusumaatmaja}}, \ and\ \bibinfo {author} {\bibfnamefont {X.}~\bibnamefont {Wang}},\ }\bibfield  {title} {\enquote {\bibinfo {title} {Reconfiguration of multistable 3d ferromagnetic mesostructures guided by energy landscape surveys},}\ }\href@noop {} {\bibfield  {journal} {\bibinfo  {journal} {Extreme Mech. Lett.}\ }\textbf {\bibinfo {volume} {48}},\ \bibinfo {pages} {101428} (\bibinfo {year} {2021})}\BibitemShut {NoStop}%
\bibitem [{\citenamefont {Yan}, \citenamefont {Yu},\ and\ \citenamefont {Mehta}(2019)}]{yan2019analytical}%
  \BibitemOpen
  \bibfield  {author} {\bibinfo {author} {\bibfnamefont {W.}~\bibnamefont {Yan}}, \bibinfo {author} {\bibfnamefont {Y.}~\bibnamefont {Yu}}, \ and\ \bibinfo {author} {\bibfnamefont {A.}~\bibnamefont {Mehta}},\ }\bibfield  {title} {\enquote {\bibinfo {title} {Analytical modeling for rapid design of bistable buckled beams},}\ }\href@noop {} {\bibfield  {journal} {\bibinfo  {journal} {Theor. Appl. Mech. Lett.}\ }\textbf {\bibinfo {volume} {9}},\ \bibinfo {pages} {264--272} (\bibinfo {year} {2019})}\BibitemShut {NoStop}%
\bibitem [{\citenamefont {Camescasse}, \citenamefont {Fernandes},\ and\ \citenamefont {Pouget}(2013)}]{camescasse2013bistable}%
  \BibitemOpen
  \bibfield  {author} {\bibinfo {author} {\bibfnamefont {B.}~\bibnamefont {Camescasse}}, \bibinfo {author} {\bibfnamefont {A.}~\bibnamefont {Fernandes}}, \ and\ \bibinfo {author} {\bibfnamefont {J.}~\bibnamefont {Pouget}},\ }\bibfield  {title} {\enquote {\bibinfo {title} {Bistable buckled beam: Elastica modeling and analysis of static actuation},}\ }\href@noop {} {\bibfield  {journal} {\bibinfo  {journal} {Int. J. Solids Struct.}\ }\textbf {\bibinfo {volume} {50}},\ \bibinfo {pages} {2881--2893} (\bibinfo {year} {2013})}\BibitemShut {NoStop}%
\bibitem [{\citenamefont {Cleary}\ and\ \citenamefont {Su}(2015)}]{cleary2015modeling}%
  \BibitemOpen
  \bibfield  {author} {\bibinfo {author} {\bibfnamefont {J.}~\bibnamefont {Cleary}}\ and\ \bibinfo {author} {\bibfnamefont {H.-J.}\ \bibnamefont {Su}},\ }\bibfield  {title} {\enquote {\bibinfo {title} {Modeling and experimental validation of actuating a bistable buckled beam via moment input},}\ }\href@noop {} {\bibfield  {journal} {\bibinfo  {journal} {J. Appl. Mech.}\ }\textbf {\bibinfo {volume} {82}},\ \bibinfo {pages} {051005} (\bibinfo {year} {2015})}\BibitemShut {NoStop}%
\bibitem [{\citenamefont {Liang}\ \emph {et~al.}(2023)\citenamefont {Liang}, \citenamefont {Wang}, \citenamefont {Luo}, \citenamefont {Takezawa}, \citenamefont {Zhang},\ and\ \citenamefont {Kang}}]{liang2023programmable}%
  \BibitemOpen
  \bibfield  {author} {\bibinfo {author} {\bibfnamefont {K.}~\bibnamefont {Liang}}, \bibinfo {author} {\bibfnamefont {Y.}~\bibnamefont {Wang}}, \bibinfo {author} {\bibfnamefont {Y.}~\bibnamefont {Luo}}, \bibinfo {author} {\bibfnamefont {A.}~\bibnamefont {Takezawa}}, \bibinfo {author} {\bibfnamefont {X.}~\bibnamefont {Zhang}}, \ and\ \bibinfo {author} {\bibfnamefont {Z.}~\bibnamefont {Kang}},\ }\bibfield  {title} {\enquote {\bibinfo {title} {Programmable and multistable metamaterials made of precisely tailored bistable cells},}\ }\href@noop {} {\bibfield  {journal} {\bibinfo  {journal} {Mater. Des.}\ }\textbf {\bibinfo {volume} {227}},\ \bibinfo {pages} {111810} (\bibinfo {year} {2023})}\BibitemShut {NoStop}%
\bibitem [{\citenamefont {ten Wolde}\ and\ \citenamefont {Farhadi}(2024)}]{ten2024single}%
  \BibitemOpen
  \bibfield  {author} {\bibinfo {author} {\bibfnamefont {M.~A.}\ \bibnamefont {ten Wolde}}\ and\ \bibinfo {author} {\bibfnamefont {D.}~\bibnamefont {Farhadi}},\ }\bibfield  {title} {\enquote {\bibinfo {title} {A single-input state-switching building block harnessing internal instabilities},}\ }\href@noop {} {\bibfield  {journal} {\bibinfo  {journal} {Mech. Mach. Theory}\ }\textbf {\bibinfo {volume} {196}},\ \bibinfo {pages} {105626} (\bibinfo {year} {2024})}\BibitemShut {NoStop}%
\bibitem [{\citenamefont {Ghavidelnia}\ \emph {et~al.}(2023)\citenamefont {Ghavidelnia}, \citenamefont {Yin}, \citenamefont {Cao},\ and\ \citenamefont {Eberl}}]{ghavidelnia2023curly}%
  \BibitemOpen
  \bibfield  {author} {\bibinfo {author} {\bibfnamefont {N.}~\bibnamefont {Ghavidelnia}}, \bibinfo {author} {\bibfnamefont {K.}~\bibnamefont {Yin}}, \bibinfo {author} {\bibfnamefont {B.}~\bibnamefont {Cao}}, \ and\ \bibinfo {author} {\bibfnamefont {C.}~\bibnamefont {Eberl}},\ }\bibfield  {title} {\enquote {\bibinfo {title} {Curly beam with programmable bistability},}\ }\href@noop {} {\bibfield  {journal} {\bibinfo  {journal} {Mater. Des.}\ }\textbf {\bibinfo {volume} {230}},\ \bibinfo {pages} {111988} (\bibinfo {year} {2023})}\BibitemShut {NoStop}%
\bibitem [{\citenamefont {Bonthron}\ and\ \citenamefont {Tubaldi}(2024)}]{bonthron2024dynamic}%
  \BibitemOpen
  \bibfield  {author} {\bibinfo {author} {\bibfnamefont {M.}~\bibnamefont {Bonthron}}\ and\ \bibinfo {author} {\bibfnamefont {E.}~\bibnamefont {Tubaldi}},\ }\bibfield  {title} {\enquote {\bibinfo {title} {Dynamic behavior of bistable shallow arches: From intrawell to chaotic motion},}\ }\href@noop {} {\bibfield  {journal} {\bibinfo  {journal} {J. Appl. Mech.}\ }\textbf {\bibinfo {volume} {91}},\ \bibinfo {pages} {021010} (\bibinfo {year} {2024})}\BibitemShut {NoStop}%
\bibitem [{\citenamefont {Hasan}\ \emph {et~al.}(2023{\natexlab{b}})\citenamefont {Hasan}, \citenamefont {Greenwood}, \citenamefont {Parker}, \citenamefont {Kong},\ and\ \citenamefont {Wang}}]{PhysRevE.108.L022201}%
  \BibitemOpen
  \bibfield  {author} {\bibinfo {author} {\bibfnamefont {M.~N.}\ \bibnamefont {Hasan}}, \bibinfo {author} {\bibfnamefont {T.~E.}\ \bibnamefont {Greenwood}}, \bibinfo {author} {\bibfnamefont {R.~G.}\ \bibnamefont {Parker}}, \bibinfo {author} {\bibfnamefont {Y.~L.}\ \bibnamefont {Kong}}, \ and\ \bibinfo {author} {\bibfnamefont {P.}~\bibnamefont {Wang}},\ }\bibfield  {title} {\enquote {\bibinfo {title} {Fractal patterns in the parameter space of a bistable duffing oscillator},}\ }\href {\doibase 10.1103/PhysRevE.108.L022201} {\bibfield  {journal} {\bibinfo  {journal} {Phys. Rev. E}\ }\textbf {\bibinfo {volume} {108}},\ \bibinfo {pages} {L022201} (\bibinfo {year} {2023}{\natexlab{b}})}\BibitemShut {NoStop}%
\bibitem [{\citenamefont {Rouleau}\ \emph {et~al.}(2024)\citenamefont {Rouleau}, \citenamefont {Keller}, \citenamefont {Lee}, \citenamefont {Craig}, \citenamefont {Shi},\ and\ \citenamefont {Meaud}}]{rouleau2024numerical}%
  \BibitemOpen
  \bibfield  {author} {\bibinfo {author} {\bibfnamefont {M.}~\bibnamefont {Rouleau}}, \bibinfo {author} {\bibfnamefont {J.}~\bibnamefont {Keller}}, \bibinfo {author} {\bibfnamefont {J.}~\bibnamefont {Lee}}, \bibinfo {author} {\bibfnamefont {S.}~\bibnamefont {Craig}}, \bibinfo {author} {\bibfnamefont {C.}~\bibnamefont {Shi}}, \ and\ \bibinfo {author} {\bibfnamefont {J.}~\bibnamefont {Meaud}},\ }\bibfield  {title} {\enquote {\bibinfo {title} {Numerical and experimental study of impact dynamics of bistable buckled beams},}\ }\href@noop {} {\bibfield  {journal} {\bibinfo  {journal} {J. Sound Vib.}\ }\textbf {\bibinfo {volume} {576}},\ \bibinfo {pages} {118291} (\bibinfo {year} {2024})}\BibitemShut {NoStop}%
\bibitem [{\citenamefont {Paar}\ and\ \citenamefont {Pavin}(1998)}]{PhysRevE.57.1544}%
  \BibitemOpen
  \bibfield  {author} {\bibinfo {author} {\bibfnamefont {V.}~\bibnamefont {Paar}}\ and\ \bibinfo {author} {\bibfnamefont {N.}~\bibnamefont {Pavin}},\ }\bibfield  {title} {\enquote {\bibinfo {title} {Intermingled fractal arnold tongues},}\ }\href {\doibase 10.1103/PhysRevE.57.1544} {\bibfield  {journal} {\bibinfo  {journal} {Phys. Rev. E}\ }\textbf {\bibinfo {volume} {57}},\ \bibinfo {pages} {1544--1549} (\bibinfo {year} {1998})}\BibitemShut {NoStop}%
\bibitem [{\citenamefont {Amor}\ \emph {et~al.}(2023)\citenamefont {Amor}, \citenamefont {Fernandes}, \citenamefont {Pouget},\ and\ \citenamefont {Maurini}}]{amor2023nonlinear}%
  \BibitemOpen
  \bibfield  {author} {\bibinfo {author} {\bibfnamefont {A.}~\bibnamefont {Amor}}, \bibinfo {author} {\bibfnamefont {A.}~\bibnamefont {Fernandes}}, \bibinfo {author} {\bibfnamefont {J.}~\bibnamefont {Pouget}}, \ and\ \bibinfo {author} {\bibfnamefont {C.}~\bibnamefont {Maurini}},\ }\bibfield  {title} {\enquote {\bibinfo {title} {Nonlinear dynamics and snap-through regimes of a bistable buckled beam excited by an electromagnetic laplace force},}\ }\href@noop {} {\bibfield  {journal} {\bibinfo  {journal} {Eur. J. Mech. A/Solids}\ }\textbf {\bibinfo {volume} {98}},\ \bibinfo {pages} {104834} (\bibinfo {year} {2023})}\BibitemShut {NoStop}%
\bibitem [{\citenamefont {Virgin}(2000)}]{virgin2000introduction}%
  \BibitemOpen
  \bibfield  {author} {\bibinfo {author} {\bibfnamefont {L.~N.}\ \bibnamefont {Virgin}},\ }\href@noop {} {\emph {\bibinfo {title} {Introduction to experimental nonlinear dynamics: a case study in mechanical vibration}}}\ (\bibinfo  {publisher} {Cambridge Univ. Press},\ \bibinfo {year} {2000})\BibitemShut {NoStop}%
\bibitem [{\citenamefont {Moon}(2008)}]{moon2008chaotic}%
  \BibitemOpen
  \bibfield  {author} {\bibinfo {author} {\bibfnamefont {F.~C.}\ \bibnamefont {Moon}},\ }\href@noop {} {\emph {\bibinfo {title} {Chaotic and fractal dynamics: introduction for applied scientists and engineers}}}\ (\bibinfo  {publisher} {John Wiley \& Sons},\ \bibinfo {year} {2008})\BibitemShut {NoStop}%
\bibitem [{\citenamefont {Moon}(1984)}]{moon1984fractal}%
  \BibitemOpen
  \bibfield  {author} {\bibinfo {author} {\bibfnamefont {F.~C.}\ \bibnamefont {Moon}},\ }\bibfield  {title} {\enquote {\bibinfo {title} {Fractal boundary for chaos in a two-state mechanical oscillator},}\ }\href@noop {} {\bibfield  {journal} {\bibinfo  {journal} {Phys. Rev. Lett.}\ }\textbf {\bibinfo {volume} {53}},\ \bibinfo {pages} {962} (\bibinfo {year} {1984})}\BibitemShut {NoStop}%
\bibitem [{\citenamefont {Xu}\ and\ \citenamefont {Xiang}(2024)}]{xu2024chaotic}%
  \BibitemOpen
  \bibfield  {author} {\bibinfo {author} {\bibfnamefont {L.}~\bibnamefont {Xu}}\ and\ \bibinfo {author} {\bibfnamefont {Z.}~\bibnamefont {Xiang}},\ }\bibfield  {title} {\enquote {\bibinfo {title} {Chaotic metastructures for frequency self-conversion},}\ }\href@noop {} {\bibfield  {journal} {\bibinfo  {journal} {Mech. Syst. Signal Process.}\ }\textbf {\bibinfo {volume} {206}},\ \bibinfo {pages} {110927} (\bibinfo {year} {2024})}\BibitemShut {NoStop}%
\bibitem [{\citenamefont {Kovacic}, \citenamefont {Brennan},\ and\ \citenamefont {Lineton}(2008)}]{kovacic2008resonance}%
  \BibitemOpen
  \bibfield  {author} {\bibinfo {author} {\bibfnamefont {I.}~\bibnamefont {Kovacic}}, \bibinfo {author} {\bibfnamefont {M.~J.}\ \bibnamefont {Brennan}}, \ and\ \bibinfo {author} {\bibfnamefont {B.}~\bibnamefont {Lineton}},\ }\bibfield  {title} {\enquote {\bibinfo {title} {On the resonance response of an asymmetric duffing oscillator},}\ }\href@noop {} {\bibfield  {journal} {\bibinfo  {journal} {Int. J. Non-Linear Mech.}\ }\textbf {\bibinfo {volume} {43}},\ \bibinfo {pages} {858--867} (\bibinfo {year} {2008})}\BibitemShut {NoStop}%
\bibitem [{\citenamefont {Sadeghi}\ and\ \citenamefont {Li}(2020)}]{sadeghi2020dynamic}%
  \BibitemOpen
  \bibfield  {author} {\bibinfo {author} {\bibfnamefont {S.}~\bibnamefont {Sadeghi}}\ and\ \bibinfo {author} {\bibfnamefont {S.}~\bibnamefont {Li}},\ }\bibfield  {title} {\enquote {\bibinfo {title} {Dynamic folding of origami by exploiting asymmetric bi-stability},}\ }\href@noop {} {\bibfield  {journal} {\bibinfo  {journal} {Extreme Mech. Lett.}\ }\textbf {\bibinfo {volume} {40}},\ \bibinfo {pages} {100958} (\bibinfo {year} {2020})}\BibitemShut {NoStop}%
\bibitem [{\citenamefont {Simo}\ and\ \citenamefont {Woafo}(2016)}]{simo2016effects}%
  \BibitemOpen
  \bibfield  {author} {\bibinfo {author} {\bibfnamefont {H.}~\bibnamefont {Simo}}\ and\ \bibinfo {author} {\bibfnamefont {P.}~\bibnamefont {Woafo}},\ }\bibfield  {title} {\enquote {\bibinfo {title} {Effects of asymmetric potentials on bursting oscillations in duffing oscillator},}\ }\href@noop {} {\bibfield  {journal} {\bibinfo  {journal} {Optik}\ }\textbf {\bibinfo {volume} {127}},\ \bibinfo {pages} {8760--8766} (\bibinfo {year} {2016})}\BibitemShut {NoStop}%
\bibitem [{\citenamefont {Abbasi}\ \emph {et~al.}(2023)\citenamefont {Abbasi}, \citenamefont {Sano}, \citenamefont {Yan},\ and\ \citenamefont {Reis}}]{abbasi2023snap}%
  \BibitemOpen
  \bibfield  {author} {\bibinfo {author} {\bibfnamefont {A.}~\bibnamefont {Abbasi}}, \bibinfo {author} {\bibfnamefont {T.~G.}\ \bibnamefont {Sano}}, \bibinfo {author} {\bibfnamefont {D.}~\bibnamefont {Yan}}, \ and\ \bibinfo {author} {\bibfnamefont {P.~M.}\ \bibnamefont {Reis}},\ }\bibfield  {title} {\enquote {\bibinfo {title} {Snap buckling of bistable beams under combined mechanical and magnetic loading},}\ }\href@noop {} {\bibfield  {journal} {\bibinfo  {journal} {Phil. Trans. R. Soc. A}\ }\textbf {\bibinfo {volume} {381}},\ \bibinfo {pages} {20220029} (\bibinfo {year} {2023})}\BibitemShut {NoStop}%
\bibitem [{\citenamefont {Pal}\ and\ \citenamefont {Sitti}(2023)}]{pal2023programmable}%
  \BibitemOpen
  \bibfield  {author} {\bibinfo {author} {\bibfnamefont {A.}~\bibnamefont {Pal}}\ and\ \bibinfo {author} {\bibfnamefont {M.}~\bibnamefont {Sitti}},\ }\bibfield  {title} {\enquote {\bibinfo {title} {Programmable mechanical devices through magnetically tunable bistable elements},}\ }\href@noop {} {\bibfield  {journal} {\bibinfo  {journal} {Proc. Natl. Acad. Sci. U.S.A.}\ }\textbf {\bibinfo {volume} {120}},\ \bibinfo {pages} {e2212489120} (\bibinfo {year} {2023})}\BibitemShut {NoStop}%
\bibitem [{\citenamefont {Zhao}, \citenamefont {Wang},\ and\ \citenamefont {Zhang}(2023)}]{zhao2023tuning}%
  \BibitemOpen
  \bibfield  {author} {\bibinfo {author} {\bibfnamefont {Z.}~\bibnamefont {Zhao}}, \bibinfo {author} {\bibfnamefont {C.}~\bibnamefont {Wang}}, \ and\ \bibinfo {author} {\bibfnamefont {X.~S.}\ \bibnamefont {Zhang}},\ }\bibfield  {title} {\enquote {\bibinfo {title} {Tuning buckling behaviors in magnetically active structures: topology optimization and experimental validation},}\ }\href@noop {} {\bibfield  {journal} {\bibinfo  {journal} {J. Appl. Mech.}\ ,\ \bibinfo {pages} {1--18}} (\bibinfo {year} {2023})}\BibitemShut {NoStop}%
\bibitem [{\citenamefont {Zhao}\ and\ \citenamefont {Zhang}(2023)}]{zhao2023encoding}%
  \BibitemOpen
  \bibfield  {author} {\bibinfo {author} {\bibfnamefont {Z.}~\bibnamefont {Zhao}}\ and\ \bibinfo {author} {\bibfnamefont {X.~S.}\ \bibnamefont {Zhang}},\ }\bibfield  {title} {\enquote {\bibinfo {title} {Encoding reprogrammable properties into magneto-mechanical materials via topology optimization},}\ }\href@noop {} {\bibfield  {journal} {\bibinfo  {journal} {npj Comput. Mater.}\ }\textbf {\bibinfo {volume} {9}},\ \bibinfo {pages} {57} (\bibinfo {year} {2023})}\BibitemShut {NoStop}%
\bibitem [{\citenamefont {Timoshenko}\ and\ \citenamefont {Gere}(2009)}]{timoshenko2009theory}%
  \BibitemOpen
  \bibfield  {author} {\bibinfo {author} {\bibfnamefont {S.~P.}\ \bibnamefont {Timoshenko}}\ and\ \bibinfo {author} {\bibfnamefont {J.~M.}\ \bibnamefont {Gere}},\ }\href@noop {} {\emph {\bibinfo {title} {Theory of elastic stability}}}\ (\bibinfo  {publisher} {Courier Corporation},\ \bibinfo {year} {2009})\BibitemShut {NoStop}%
\bibitem [{\citenamefont {Tseng}\ and\ \citenamefont {Dugundji}(1971)}]{tseng1971nonlinear}%
  \BibitemOpen
  \bibfield  {author} {\bibinfo {author} {\bibfnamefont {W.-Y.}\ \bibnamefont {Tseng}}\ and\ \bibinfo {author} {\bibfnamefont {J.}~\bibnamefont {Dugundji}},\ }\bibfield  {title} {\enquote {\bibinfo {title} {Nonlinear vibrations of a buckled beam under harmonic excitation},}\ }\href@noop {} {\bibfield  {journal} {\bibinfo  {journal} {J. Appl. Mech.}\ } (\bibinfo {year} {1971})}\BibitemShut {NoStop}%
\bibitem [{SI()}]{SI}%
  \BibitemOpen
  \href@noop {} {\enquote {\bibinfo {title} {See supplemental information at url for additional results, detailed derivations, and calculation procedures.}}\ }\BibitemShut {NoStop}%
\bibitem [{smo(7 16)}]{smooth-on}%
  \BibitemOpen
  \href@noop {} {\enquote {\bibinfo {title} {Dragon skin 30},}\ }\bibinfo {howpublished} {\url{https://www.smooth-on.com/products/dragon-skin-30/}} (\bibinfo {year} {Accessed 2023-07-16})\BibitemShut {NoStop}%
\bibitem [{\citenamefont {Ranzani}\ \emph {et~al.}(2015)\citenamefont {Ranzani}, \citenamefont {Gerboni}, \citenamefont {Cianchetti},\ and\ \citenamefont {Menciassi}}]{ranzani2015bioinspired}%
  \BibitemOpen
  \bibfield  {author} {\bibinfo {author} {\bibfnamefont {T.}~\bibnamefont {Ranzani}}, \bibinfo {author} {\bibfnamefont {G.}~\bibnamefont {Gerboni}}, \bibinfo {author} {\bibfnamefont {M.}~\bibnamefont {Cianchetti}}, \ and\ \bibinfo {author} {\bibfnamefont {A.}~\bibnamefont {Menciassi}},\ }\bibfield  {title} {\enquote {\bibinfo {title} {A bioinspired soft manipulator for minimally invasive surgery},}\ }\href@noop {} {\bibfield  {journal} {\bibinfo  {journal} {Bioinspir. Biomim.}\ }\textbf {\bibinfo {volume} {10}},\ \bibinfo {pages} {035008} (\bibinfo {year} {2015})}\BibitemShut {NoStop}%
\end{thebibliography}%
\end{document}

\preprint{AIP/123-QED}

\title{Supplementary Materials for\\Harmonically Induced Shape Morphing of Bistable Buckled Beam with Static Bias}
\author{Md Nahid Hasan}
\affiliation{Department of Mechanical Engineering, University of Utah, Salt Lake City, UT 84112, USA}
\affiliation{Department of Mechanical Engineering, Montana Technological University, Butte, MT 59701, USA}

\author{Sharat Paul}%
\affiliation{Department of Mechanical Engineering, University of Utah, Salt Lake City, UT 84112, USA}

\author{Taylor E. Greenwood}%
\affiliation{Department of Mechanical Engineering, University of Utah, Salt Lake City, UT 84112, USA}
\affiliation{Department of Mechanical Engineering, Pennsylvania State University, University Park, PA 16802, USA}

\author{Robert G. Parker}%
\affiliation{Department of Mechanical Engineering, University of Utah, Salt Lake City, UT 84112, USA}

\author{Yong Lin Kong}%
\affiliation{Department of Mechanical Engineering, University of Utah, Salt Lake City, UT 84112, USA}
\affiliation{Department of Mechanical Engineering, Rice University, Houston, TX 77005, USA}
%\thanks{Authors to whom correspondence should be addressed: yong.kong@utah.edu}

\author{Pai Wang}%
\affiliation{Department of Mechanical Engineering, University of Utah, Salt Lake City, UT 84112, USA}
\thanks{Authors to whom correspondence should be addressed: pai.wang@utah.edu}

\date{\today}% It is always \today, today,
             %  but any date may be explicitly specified
%\begin{abstract}
%\end{abstract}

%\keywords{Suggested keywords}%Use showkeys class option if keyword
                              %display desired
\maketitle
%\newpage
\tableofcontents
\clearpage
\section{Continuous buckled beam equation to discrete bistable Duffing equation}
We consider a straight beam buckled into its first buckled mode and apply harmonic excitation to switch its state from the first stable state to the second stable state, as shown in Figs. 1(a) and 1(b) in the main manuscript. Furthermore, We consider the transverse deflection of the beam at a position \(x\) and time \(t\) as \(\hat{W}(x,t)\). The beam has a length \(L\), a uniform density \(\rho\), a cross-sectional area \(A\), and a flexural rigidity \(EI\), where \(E\) is Young's modulus and \(I = \frac{bt^3}{12}\) is the moment of inertia of the beam. Here, \(t\) and \(b\) are the thickness and the out-of-plane width of the beam, respectively (see Table I). Initially, the bistable beam is modeled using a continuous beam vibration equation, which is then discretized into a bistable Duffing equation. The governing differential equation for a buckled beam, originally flat and then compressed past its critical buckling load, results in a static displacement \(\hat{W}_{o}\) when the ends are fixed. This beam is subjected to a point load harmonic excitation (Figs. 1(a) and 1(b) in the main manuscript). Equation~\ref{eq:S1} shows the Euler-Bernoulli beam equation of a buckled beam.
\begin{equation}\label{eq:S1}
\begin{aligned}
\ EI \frac{\partial^4 \hat W}{\partial \hat x^4}+\hat P_{cr}\frac{\partial^2 \hat W}{\partial \hat x^2}+  \rho A \frac{\partial^2 \hat W}{\partial \hat t^2} +\hat C^d \frac{\partial \hat W}{\partial \hat t}- \left\{{ \frac{EA}{2L} \int_0^L\left[\left(\frac{\partial \hat W}{\partial \hat x}\right)^2+2\frac{\partial \hat W}{\partial \hat x}\frac{\partial \hat W_o}{\partial \hat x}\right] {d\hat x}}\right\} \\ \left(\frac{\partial^2 \hat W}{\partial\hat x^2}+\frac{\partial^2\hat W_o}{\partial\hat x^2}\right) = \hat F \cos{(\hat \Omega \hat t)}.
\end{aligned}
\end{equation}
Equation\,\eqref{eq:S1} can be nondimensionalized using the nondimensionalized parameter from the Table I,
\begin{equation}\label{eq:S2}
\begin{aligned}
    \frac{\partial^4 W}{\partial x^4} &= \frac{\partial^4 (\hat Wh)}{\partial (\hat xL)^4} = \frac{h}{L^4}\frac{\partial^4 \hat W}{\partial \hat x^4}; \quad
    \frac{\partial^2 W}{\partial x^2} = \frac{\partial^2 (\hat Wh)}{\partial (\hat xL)^2} = \frac{h}{L^2}\frac{\partial^2 \hat W}{\partial \hat x^2};\quad
     \frac{\partial W}{\partial x} = \frac{\partial (\hat Wh)}{\partial (\hat xL)} = \frac{h}{L}\frac{\partial \hat W}{\partial \hat x};\\
     \left(\frac{\partial W_o}{\partial x}\right)^2 &= \frac{h^2}{L^2}\left(\frac{\partial \hat W_o}{\partial \hat x}\right)^2; \quad
     \left(\frac{\partial W_o}{\partial x}\right) = \frac{h}{L}\left(\frac{\partial \hat W_o}{\partial \hat x}\right); \quad
     \frac{\partial^2 W_o}{\partial x^2} = \frac{\partial^2 (\hat W_oh)}{\partial (\hat xL)^2} = \frac{h}{L^2}\frac{\partial^2 \hat W_o}{\partial \hat x^2};\\
     \frac{\partial W}{\partial t} &= \frac{h}{T}\frac{\partial \hat W}{\partial \hat t};\quad
     \frac{\partial^2 W}{\partial t^2} = \frac{\partial^2 (\hat Wh)}{\partial (\hat tT)^2} = \frac{h}{T^2}\frac{\partial^2 \hat W}{\partial \hat t^2};\quad
     dx = d(\hat xL) = L d\hat x,
\end{aligned}  
\end{equation}
where $\hat W_o(x=\frac{L}{2})= h$ is the apex height or amplitude of the bistable buckled beam and replace the above substitution in Eq.\,\eqref{eq:S1} we get,
\begin{equation}\label{eq:S3}
  \begin{aligned}
\ \frac{EIh}{L^4} \frac{\partial^4 W}{\partial x^4}+\frac{\hat P_{cr}h}{L^2} \frac{\partial^2 W}{\partial x^2}+ \frac{\rho A h}{T^2}  \frac{\partial^2 W}{\partial t^2} +\frac{\hat C^d h}{T}\frac{\partial W}{\partial t}- \left\{{\frac{EAh^3}{2L^4} \int_0^L\left[\left(\frac{\partial W}{\partial x}\right)^2+2\frac{\partial W}{\partial  x}\frac{\partial W_o}{\partial x}\right] {dx}}\right\}\\ \left(\frac{\partial^2 W}{\partial x^2}+\frac{\partial^2 W_o}{\partial x^2}\right)&\\ =  \hat F \cos({\hat \Omega T t}).
\end{aligned}
\end{equation}

Now divide both sides of the Eq.\,\eqref{eq:S3} by, $\frac{EIh}{L^4}$
\begin{equation}\label{eq:S4}
  \begin{aligned}
\ \frac{\partial^4 W}{\partial x^4}+\frac{\hat P_{cr}L^2}{EI} \frac{\partial^2 W}{\partial x^2}+ \frac{\rho A L^4}{EIT^2}  \frac{\partial^2 W}{\partial t^2} +\frac{\hat C^d L^4}{EIT}\frac{\partial W}{\partial t}- \left\{{\frac{Ah^2}{2I} \int_0^L\left[\left(\frac{\partial W}{\partial x}\right)^2+2\frac{\partial W}{\partial  x}\frac{\partial W_o}{\partial x}\right] {dx}}\right\}\\ \left(\frac{\partial^2 W}{\partial x^2}+\frac{\partial^2 W_o}{\partial x^2}\right)&\\ = \frac{\hat F L^4}{EIh} \cos({\hat \Omega T t}),
\end{aligned}
\end{equation}
Next, T is defined by setting the coefficient of the inertia term equal to unity,
\begin{equation}\label{eq:S5}
    \frac{\rho A L^4}{EIT^2}=1 \implies T=\sqrt{\frac{\rho A L^4}{EI}},
\end{equation}
$T$ is called the time constant. Using the nondimensional substitutions from Table II and Eq.\,\eqref{eq:S5}, Eq.\,\eqref{eq:S4} becomes Eq.\,\eqref{eq:S6}.
\begin{equation}\label{eq:S6}
  \begin{aligned}
\ \frac{\partial^4 W}{\partial x^4}+P_{cr}\frac{\partial^2 W}{\partial x^2}+ \frac{\partial^2 W}{\partial t^2} +C^d \frac{\partial W}{\partial t}- \left\{{6Q^2 \int_0^L\left[\left(\frac{\partial W}{\partial x}\right)^2+2\frac{\partial W}{\partial  x}\frac{\partial W_o}{\partial x}\right] {dx}}\right\}\\ \left(\frac{\partial^2 W}{\partial x^2}+\frac{\partial^2 W_o}{\partial x^2}\right)&\\ = F \cos({\hat \Omega T t}).
\end{aligned}
\end{equation}

To solve  Eq.\,\eqref{eq:S6}, we apply the separation-of-variables method, considering a solution as the product of a spatial function $\phi_i(x)$ and a time-dependent function $q_i(t)$. Using Galerkin's method, Eq.\,\eqref{eq:S6} is transformed into a set of coupled ordinary differential equations (ODEs). With $n$ representing the degrees of freedom (DOF), the separation of variables on Eq.\,\eqref{eq:S6} leads to:
\begin{equation}\label{eq:S7}
W(x,t) = \sum_{i=1}^{n} q_i(t) \phi_i(x).
\end{equation}
% Begin the centered table environment
\begin{table}[h!]
\centering
\caption{Beam Geometry and Material Properties (Dragon Skin 30)}
\begin{tabular}{ |p{6cm}|p{4cm}|p{3cm}| }
%\hline
%\multicolumn{3}{|c|}{Beam Geometry and Material Properties (Dragon Skin 30)} \\
\hline
\centering Parameter & Symbol & Value \\
\hline
Beam length (mm) & $L$ & 60 \\
Beam width (mm) & $b$ & 10 \\
Beam thickness (mm) & $t$ & 1 \\
Buckled height (mm) & $h$ & 5.22 \\
Modulus of elasticity $(\textrm{MPa})$ & $E$ & $0.74\pm 0.07$ \\
Density $(\frac{kg}{m^3})$ & $\rho$ & 1082 \\
\hline
\end{tabular}
\end{table}
\newline Mode shape of the bistable buckled beam is given by,

Odd mode, 
\begin{equation}\label{eq:S8}
\\ \phi_i(x)=\frac{1}{2}\left[1-\cos (N_i x)\right] ,
\end{equation}

\begin{equation}\label{eq:S9}
\\ N_i=(i+1)\pi
\end{equation}
Here, $i=1,3,5,......$

and even modes,

\begin{equation} \label{eq:S10}
\\ \phi_i(x)=\frac{1}{2}\left[1-2x-\cos (N_ix) + \frac{2\sin (N_ix)}{N_i}\right],
\end{equation}

\begin{equation}\label{eq:S11}
\\ N_i=2.86\pi,4.92\pi....
\end{equation}

Here, $i=2,4,6,......$
%Here, $\phi_i(x)$ are the mode shapes of the beam, and $q_i(t)$ denote the generalized time-dependent coordinates. 
%\newline
\begin{table}
\caption{Nondimentional substitution}
\centering
\begin{tabular}{ |p{6cm}|p{3cm}|p{3cm}|}
%\multicolumn{2}{|c|}{Nondimensional Substitutions} \\
\hline
\centering
Parameter & Substitutions \\
\hline
$x-$ direction position    &    $x=\frac {\hat x}{L}$ \\
$z-$ direction position    &    $w=\frac {\hat w}{h}$ \\
Time & $t=\frac {\hat t}{T}$ \\
Damping Coefficient & $C^d=\frac{\hat C^d L^2}{\sqrt{\rho AEI}}$ \\
Axial Load & $P_{cr}=\hat P_{cr}\frac{L^2}{EI}$\\
Force constant & $r=\frac{\hat F L^4}{EIh} $\\
Geometric parameter & $Q=\left(\frac{h}{t}\right)$ \\
Time constant & $T=\sqrt{\frac{\rho AL^4}{EI}}$ \\
\hline
\end{tabular}\\\
\end{table}
Now plugging Eq.\,\eqref{eq:S7} into the  Eq.\,\eqref{eq:S6}, which yields a coupled set of $n$ ODE’s for $q_i$.
\begin{equation} \label{eq:S12}
\begin{split}
\sum_{i=1}^{n}\phi_i\frac{\partial^2 q_i}{\partial t^2} + C^d\sum_{i=1}^{n}\phi_i\frac{\partial q_i}{\partial t}+P_{cr}\sum_{i=1}^{n}\frac{\partial^2\phi_i}{\partial x^2}q_i +\sum_{i=1}^{n}\frac{\partial^4\phi_i}{\partial x^4}q_i \\-\left\{6Q^2 \int_0^L \left[ \left(\frac{\partial \phi_i}{\partial x}\right)^2q^2_i + 2 \left(\frac{\partial \phi_i}{\partial x}\right)\left(\frac{d\phi_o}{d x}\right)q_i\right] dx\right\}
 \left(\frac{\partial^2 \phi_i}{\partial x^2}q_i+\frac{d^2 \phi_o}{d x^2}\right)=F\cos{(\hat\Omega T\hat t)}.
\end{split}
\end{equation}
As the buckling mode shapes are orthogonal, the linear terms in the  Eq.\,\eqref{eq:S12} can be decoupled by multiplying through $\phi_j$ and integrating over the length of the beam. This provides a set of ordinary differential equations,
\begin{equation}\label{eq:S13}
\begin{split}
M_{i}{\ddot q_i} + C^d M_{i}{\dot q_i} + P_{cr} E_{i}q_i + N_{i}q_i - 6Q^2\left[(D_i q_i^2 + 2G_i q_i)\right] (E_{i} q_i + H_{i}) \
= F_i F\cos{(\hat{\Omega} Tt)},
\end{split}
\end{equation}
%Here $i=1,2,3......n$ \newline
where
\begin{equation}\label{eq:S14} 
\begin{split}
M_{i} &= \int_0^1 \phi_j \phi_i \, dx; \quad N_{i} = \int_0^1 \phi_j \frac{d^4 \phi_i}{dx^4} \, dx; \quad D_{i} = \int_0^1 \left( \frac{d \phi_i}{dx} \right)^2 \, dx; \\
E_{i} &= \int_0^1 \phi_j \frac{d^2 \phi_i}{dx^2} \, dx; \quad
F_{i} = \int_0^1 \phi_j \, dx; \quad
G_i = \int_0^1 \left( \frac{d \phi_i}{dx} \right) \left( \frac{d \phi_o}{dx} \right) \, dx; \quad
H_i = \int_0^1 \phi_j \frac{d^2 \phi_o}{dx^2} \, dx.
\end{split}
\end{equation}

Mode shapes $\phi_i$ and $\phi_j$ are orthogonal to each other, a key concept in structural dynamics. The orthonormality condition for these mode shapes is defined as follows:

\begin{equation}\label{eq:S15} 
\int_0^1 \phi_i \phi_j \, dx = \delta_{ij},
\end{equation}
The Kronecker delta, $\delta_{ij}$, signifies the orthonormality and is defined as:
\begin{equation}\label{eq:S16} 
\delta_{ij} =
\begin{cases}
1, & \text{if } i = j \\
0, & \text{if } i \neq j,
\end{cases}
\end{equation}
This orthonormality condition ensures that the integral of the product of two different mode shapes, $\phi_i$ and $\phi_j$, over their domain is zero when $i \neq j$, and is equal to 1 when $i = j$.

If we consider the first buckled mode approximation of Eq.\,\eqref{eq:S8},
\begin{equation}\label{eq:S17}
\\ \phi_1(x)=\frac{1}{2}\left[1-\cos (2\pi x)\right],
\end{equation}
Therefore, first Buckling mode parameters,
\begin{equation}\label{eq:S18}
\begin{split}
M_1=\int_0^1\phi_1 \phi_1=0.3750; \quad
N_{1}=\int_0^1\phi_1 \frac{d^4\phi_1}{dx^4}=194.8182;\\
D_{1}=\int_0^1\left(\frac{d \phi_1}{dx}\right)^2=4.9348;\quad
E_{1}=\int_0^1\phi_1 \frac{d^2\phi_1}{d x^2}=-4.9348; \quad
F_{1}=\int_0^1\phi_1=0.500;\\
G_1=\int_0^1 \left(\frac{d \phi_1}{dx}\right)\left(\frac{d\phi_o}{dx}\right)=4.9348;\quad
H_1=\int_0^1\phi_1 \frac{d^2 \phi_o}{dx^2}=-4.9348,
\end{split}
\end{equation}
Therefore, Eq.\,\eqref{eq:S13} becomes,
\begin{equation}\label{eq:S19}
\begin{split}
M_{1}{\ddot q_1} + C^d M_{1}{\dot q_1} + P_{cr} E_{1}q_1 + N_{1}q_1 - 6Q^2\left[(D_1 q_1^2 + 2G_1 q_1)\right] (E_{1} q_1 + H_{1}) \\
= F_1 F\cos{(\hat{\Omega} Tt)},
\end{split}
\end{equation}
\begin{equation}\label{eq:S20}
\begin{split}
M_{1}{\ddot q_1}+ C^d M_{1} {\dot q_1}+P_{cr} E_{1}q_1 +N_{1}q_1-
6Q^2\left[D_1E_1q^3_1+D_1H_1q^2_1+2E_1G_1q^2_1+2G_1H_1q_1\right]\\
= F_1 F\cos{(\hat \Omega Tt)},
\end{split}
\end{equation}
As $D_1=G_1$, and $E_1=H_1$, Eq.\,\eqref{eq:S20} simplifies, 
\begin{equation}\label{eq:S21}
\begin{split}
M_{1}{\ddot q_1}+ C^d M_{1} {\dot q_1}+P_{cr} E_{1}q_1 +N_{1}q_1-
6Q^2\left[D_1E_1q^3_1+D_1E_1q^2_1+2D_1E_1q^2_1+2D_1E_1q_1\right]\\
= F_1 F\cos{(\hat \Omega Tt)},
\end{split}
\end{equation}
\begin{equation}\label{eq:S22}
\begin{split}
M_{1}{\ddot q_1}+ C^d M_{1} {\dot q_1}+P_{cr} E_{1}q_1 +N_{1}q_1-
6Q^2\left[D_1E_1q^3_1+3D_1E_1q^2_1+2D_1E_1q_1\right]\\
= F_1 F\cos{(\hat \Omega Tt)},
\end{split}
\end{equation}
As $(N_1 + P_{cr} E_1) = (194.8182 - 194.8181) \approx 0$,
\begin{equation}\label{eq:S23}
M_{1} \ddot{q}_1 + C^d M_{1} \dot{q}_1 - 
6Q^2 \left[D_1 E_1 q^3_1 + 3D_1 E_1 q^2_1 + 2 D_1 E_1 q_1 \right]
= F_1 F \cos{(\hat{\Omega} T t)},
\end{equation}
and let, $K=6Q^2D_1E_1$ and $Q=h/t$, so the final form of the Eq.\,\eqref{eq:S20},
\begin{equation}\label{eq:S24}
M_{1} \ddot{q}_1 + C^d M_{1} \dot{q}_1 - 
\left[K q^3_1 + 3K q^2_1 + 2 K q_1 \right]
= F_1 F \cos{(\hat{\Omega} T t)}.
\end{equation}
We shift the equilibrium to zero for the new variable \( u \) using the transformation \( q_1 = u - 1 \). This simplifies the equations, aids in linearizing the system, and makes analysis and computations easier. It aligns with standard approximation methods, enhancing both analytical and numerical analysis. For the given equation, we apply this transformation as:

\begin{equation}\label{eq:S25}
u = q_1 + 1 \quad \implies \quad q_1 = u - 1,
\end{equation}

Using the above equation Eq.\,\eqref{eq:S24} becomes,
\begin{equation}\label{eq:S26}
M_{1}{\ddot u}+ C^d M_{1} {\dot u}-K u+Ku^3
= F_1 F\cos{(\hat \Omega Tt)},
\end{equation}

\begin{equation}\label{eq:S27}
\begin{split}
{\ddot u}+ C^d {\dot u}-\frac{K}{M_1} u_1+\frac{K}{M_1}u^3 
= F_1\left(\frac{F}{M_1}\right) \cos{(\hat \Omega Tt)},
\end{split}
\end{equation}

\begin{equation}\label{eq:S28}
\begin{split}
{\ddot u}+ C^d {\dot u}-\omega^2_{non} u+\omega^2_{non}u^3 
= F_1\left(\frac{F}{M_1}\right) \cos{(\hat \Omega Tt)}
\end{split}.
\end{equation}

$\omega_{non}$ is the first natural frequency of the first mode at the linear limit,
\begin{equation}\label{eq:S29}
\begin{split}
\omega^2_{non}=\left(\frac{K}{M_1}\right)=4\pi^4Q^2.
\end{split}
\end{equation}

%Now, the resemblance with the standard non-dimensional bistable Duffing Eq.\,\eqref{eq:S28} can be related to the equation below,
 %\begin{equation}\label{eq:S30}
    % \frac{d^2u}{d\tau^2}+\gamma\frac{du}{d\tau}-u+\alpha u^3=g\cos(\omega\tau).
 %\end{equation}

We scale the time to transform Eq.\,\eqref{eq:S28} into the standard form of a bistable Duffing equation by introducing the scaled time variable \( \tau = \omega_{\text{non}} t \), where \( \omega_{\text{non}} \) is the first natural frequency of the system at the linear limit. %This scaling aligns the temporal framework with the system's characteristic time scale, simplifying the equation and rendering it non-dimensional. 
We can write Eq.\,\eqref{eq:S28} like this,
\begin{equation}\label{eq:S30}
    \frac{d^2u}{dt^2}+C^d\frac{du}{dt}-\omega^2_{non}u+\omega^2_{non}u^3\\
=F_1\left(\frac{F}{M_1}\right)\cos{(\hat \Omega Tt)}.
\end{equation}

Let $g=F_1\left(\frac{F}{M_1}\right)$, 
\begin{equation}\label{eq:S31}
    \frac{d^2u}{dt^2}+C^d\frac{du}{dt}-\omega^2_{non}u+\omega^2_{non}u^3\\
=g \cos{(\hat \Omega Tt)},
\end{equation}

As, $\tau=\omega_{non} t$,
\begin{equation}\label{eq:S32}
    \omega^2_{non}\frac{d^2u}{d\tau^2}+\omega_{non} C^d\frac{du}{d\tau}-\omega^2_{non}u+\omega^2_{non}u^3\\
=g  \cos{\left(\frac{\hat \Omega T}{\omega_{non}}\tau\right)},
\end{equation}
dividing both sides of the above equation by $\omega^2_{non}$

\begin{equation}\label{eq:S33}
    \frac{d^2u}{d\tau^2}+\frac{C^d}{\omega_{non}} \frac{du}{d\tau}-u+u^3\\
=\frac{g}{\omega^2_{non}} \cos{\left(\frac{\hat \Omega T}{\omega_{non}}\tau\right)}.
\end{equation}
We assume the nondimensional excitation frequency as \(\omega = \frac{\hat{\Omega} T}{\omega_{\text{non}}}\). Let \( G = \frac{g}{\omega^2_{\text{non}}} \) and \(\gamma = \frac{C^d}{\omega_{\text{non}}}\). With these substitutions, Eq.\,\eqref{eq:S33} becomes:
\begin{equation}\label{eq:S34}
\begin{split}
{\ddot u} + \gamma {\dot u} - u + u^3 = G \cos{(\omega \tau)},
\end{split}
\end{equation}
where \(G\) is the point modal force amplitude (projected force with respect to the first mode shape) considering time scaling,
\begin{equation}\label{eq:S35}
\begin{split}
G &= F_1 \left( \frac{\hat{F}/L \cdot \frac{L^4}{EI \cdot h}}{4\pi^4Q^2 \cdot M_1} \right), \\
\text{with} \quad F &= \hat{F} \cdot \frac{L^3}{EIh}, \quad F_{1} = \int_0^1 \phi_1 = 0.500, \\
\text{and} \quad M_1 &= \int_0^1 \phi_1 \phi_1 = 0.3750.
\end{split}
\end{equation}
The nondimensional excitation frequency \(\omega\) (considering time scaling and the time constant \(T\)) is given by:
\begin{equation}\label{eq:S36}
\omega = \hat{\Omega} \cdot \sqrt{4 \pi^4 Q^2} \cdot \sqrt{\frac{EI}{\rho A L^4}}.
\end{equation}
\clearpage
\section{Impact of static bias force P(B) on the symmetric bistable system}
%%%%%%% Start of Figure 3 %%%%%%%
\begin{figure}[htbp]
\centering
\includegraphics[width=0.60\textwidth]{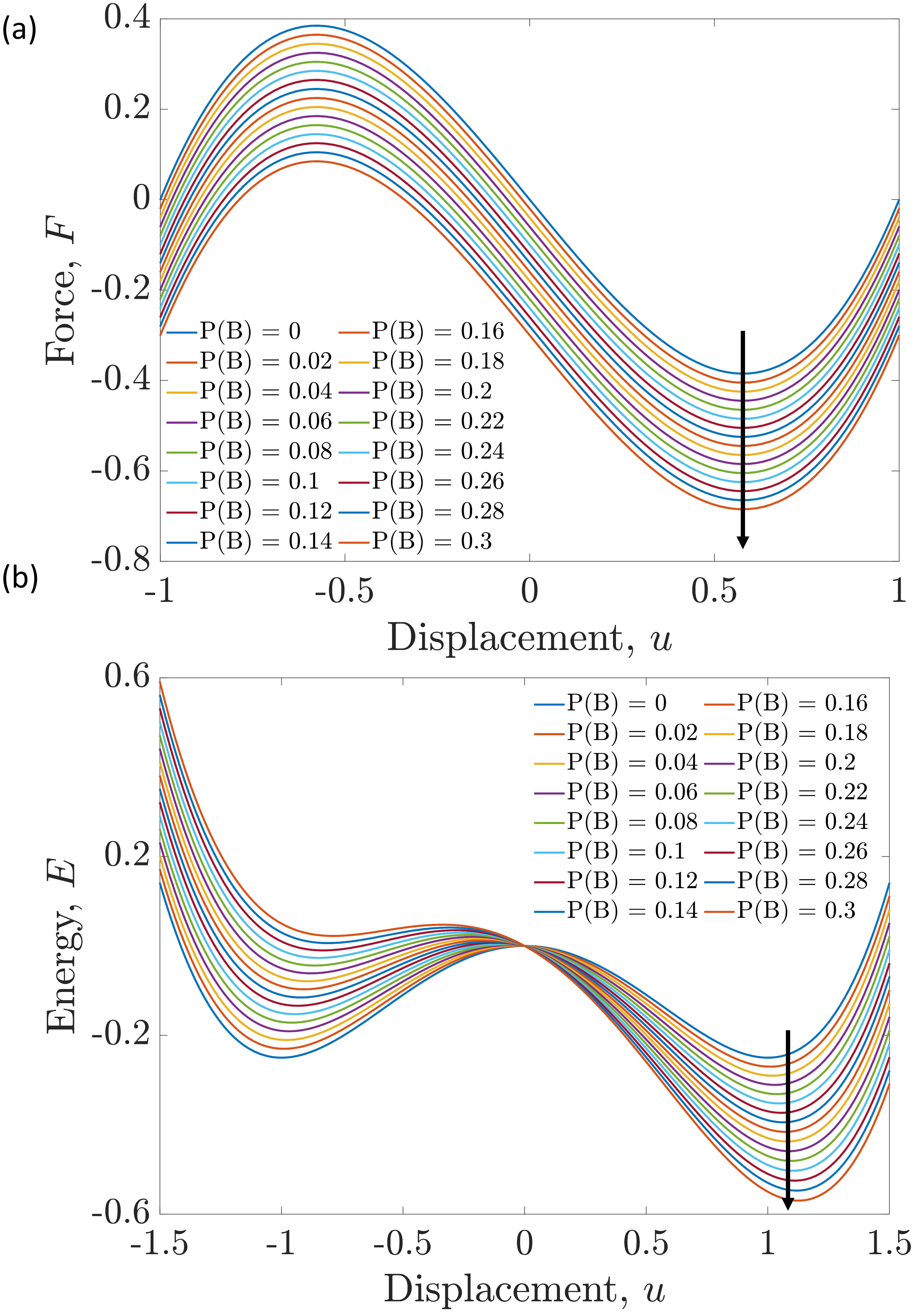}
\caption{Tunability of the bistable behavior: (a) Evolution of the force-displacement curve, demonstrating linear shifts along the force axis. (b) Linear translation of the energy landscape under various applied static bias forces. The black arrows indicate the direction of translation.}
\label{fig:F1}
\end{figure}
\clearpage
%%%%%%% Start of Figure 3 %%%%%%%
\section{Forcing amplitude-frequency parameter space for a buckled beam under combined static bias force and dynamic excitation}
\begin{figure}[htbp]
\centering
\includegraphics[width=\textwidth]{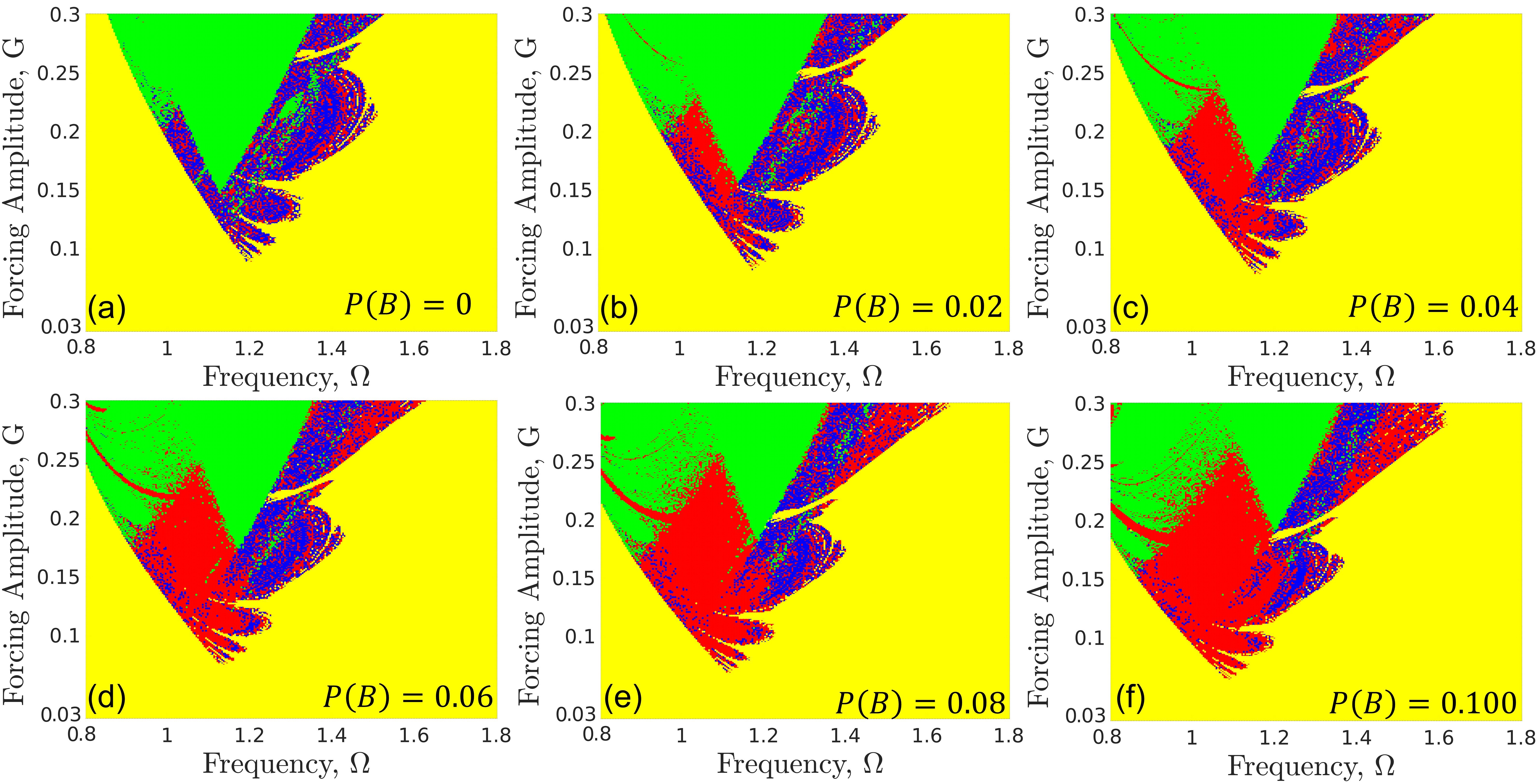}
\caption{Forcing amplitude-frequency parameter space for a bistable buckled beam with a damping ratio of \(\gamma = 0.07\) across static bias forces from \(P\textrm{(B)} = 0.00\) to \(0.100\) in increments of \(0.02\).}
\label{fig:F2}
\end{figure}
%%%%%%% End of Figure 3 %%%%%%%

%%%%%%% Start of Figure 4 %%%%%%%
\begin{figure}[htbp]
\centering
\includegraphics[width=\textwidth]{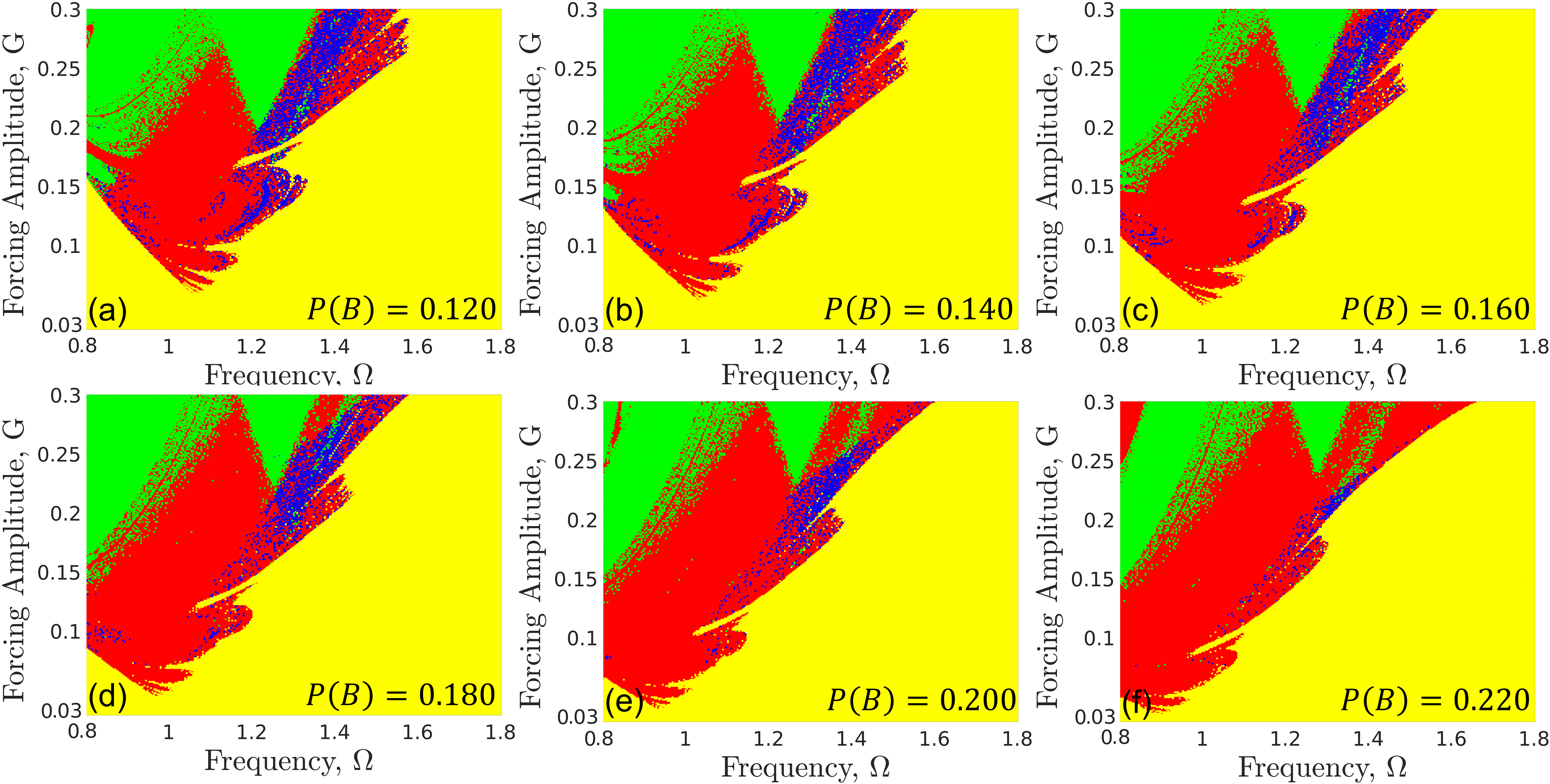}
\caption{Forcing amplitude-frequency parameter space for a bistable buckled beam with a damping ratio of \(\gamma = 0.07\) across static bias forces from \(P\textrm{(B)} = 0.120\) to \(0.200\) in increments of \(0.02\).}
\label{fig:F3}
\end{figure}
%%%%%%% End of Figure 4 %%%%%%%

%%%%%%% Start of Figure 5 %%%%%%%
\begin{figure}[htbp]
\centering
\includegraphics[width=\textwidth]{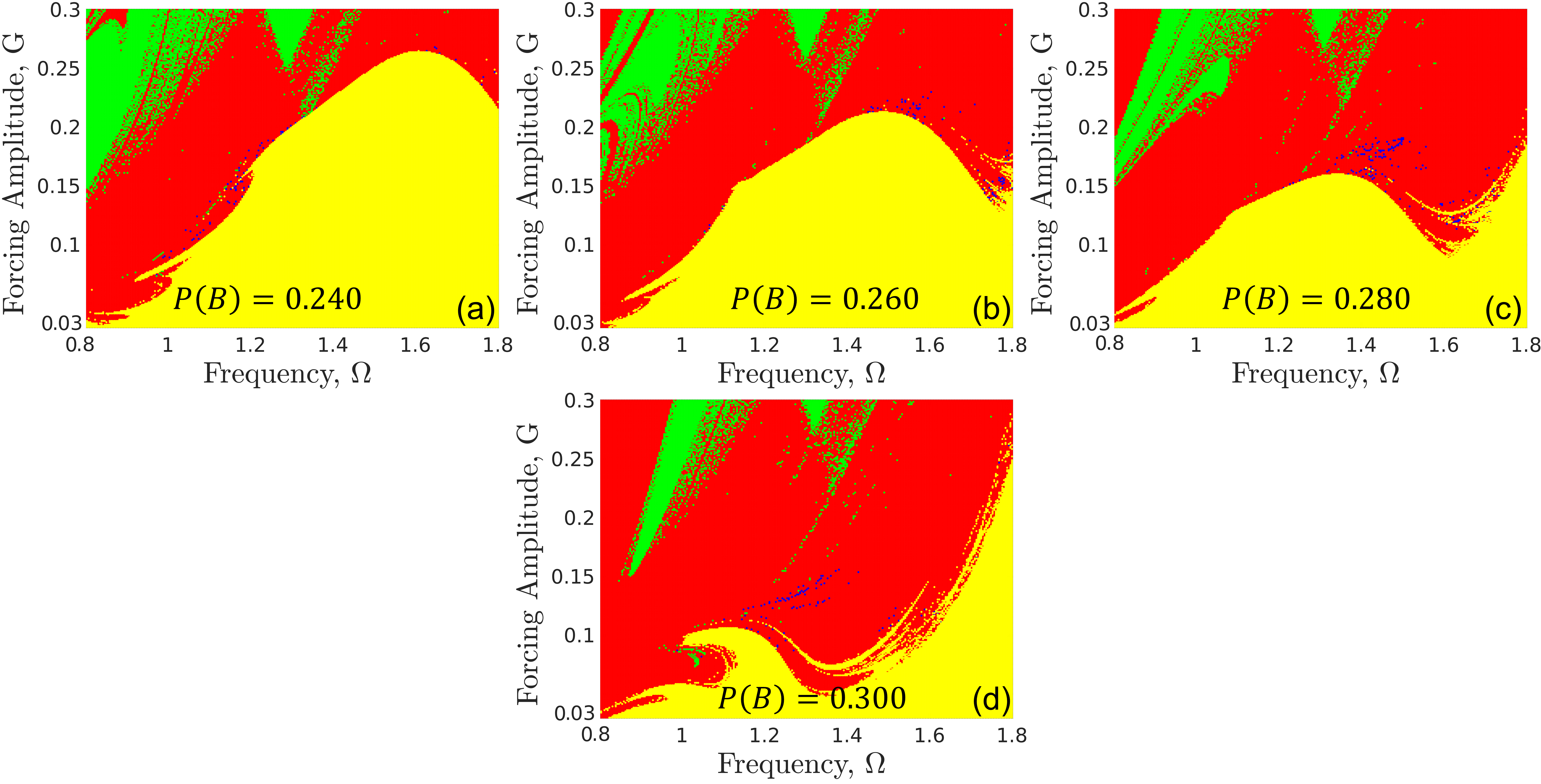}
\caption{Forcing amplitude-frequency parameter space for a bistable buckled beam with a damping ratio of \(\gamma = 0.07\) across static bias forces from \(P\textrm{(B)} = 0.240\) to \(0.300\) in increments of \(0.02\).}
\label{fig:F4}
\end{figure}
%%%%%%% End of Figure 5 %%%%%%%
\clearpage
\section{Forcing amplitude-frequency parameter space with distinct switching area}

%%%%%%% Start of Figure 5 %%%%%%%
\begin{figure}[htbp]
\centering
\includegraphics[width=\textwidth]{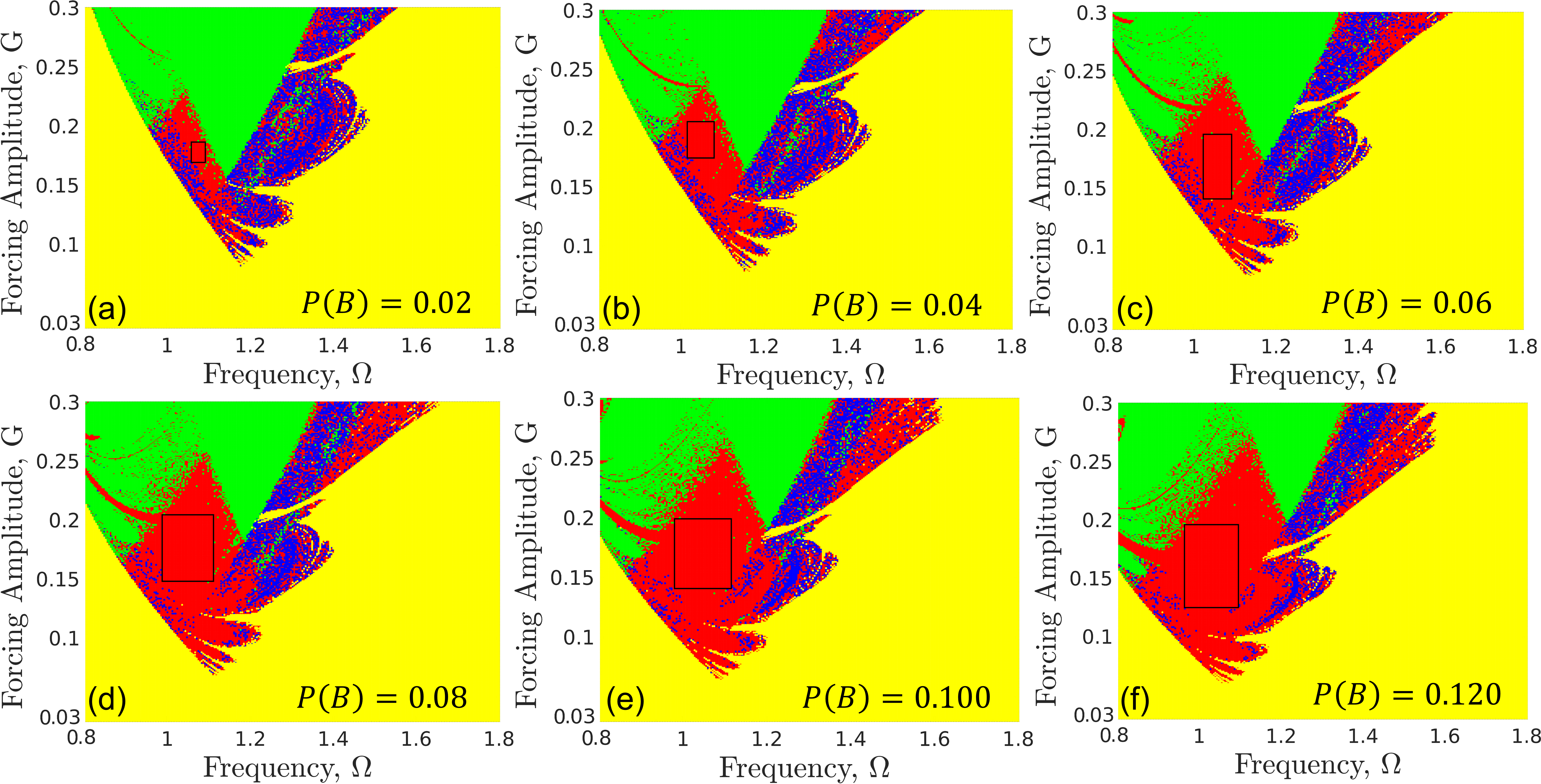}
\caption{Forcing amplitude-frequency parameter space for a bistable buckled beam with a damping ratio of \(\gamma = 0.07\) across static bias forces from \(P\textrm{(B)} = 0.02\) to \(0.120\) in increments of \(0.02\). A distinct switching area is marked by a rectangle indicating where every combination of \(G\) and \(\Omega\) results in switching behavior.}
\label{fig:F5}
\end{figure}

%%%%%%% Start of Figure 6 %%%%%%%
\begin{figure}[htbp]
\centering
\includegraphics[width=\textwidth]{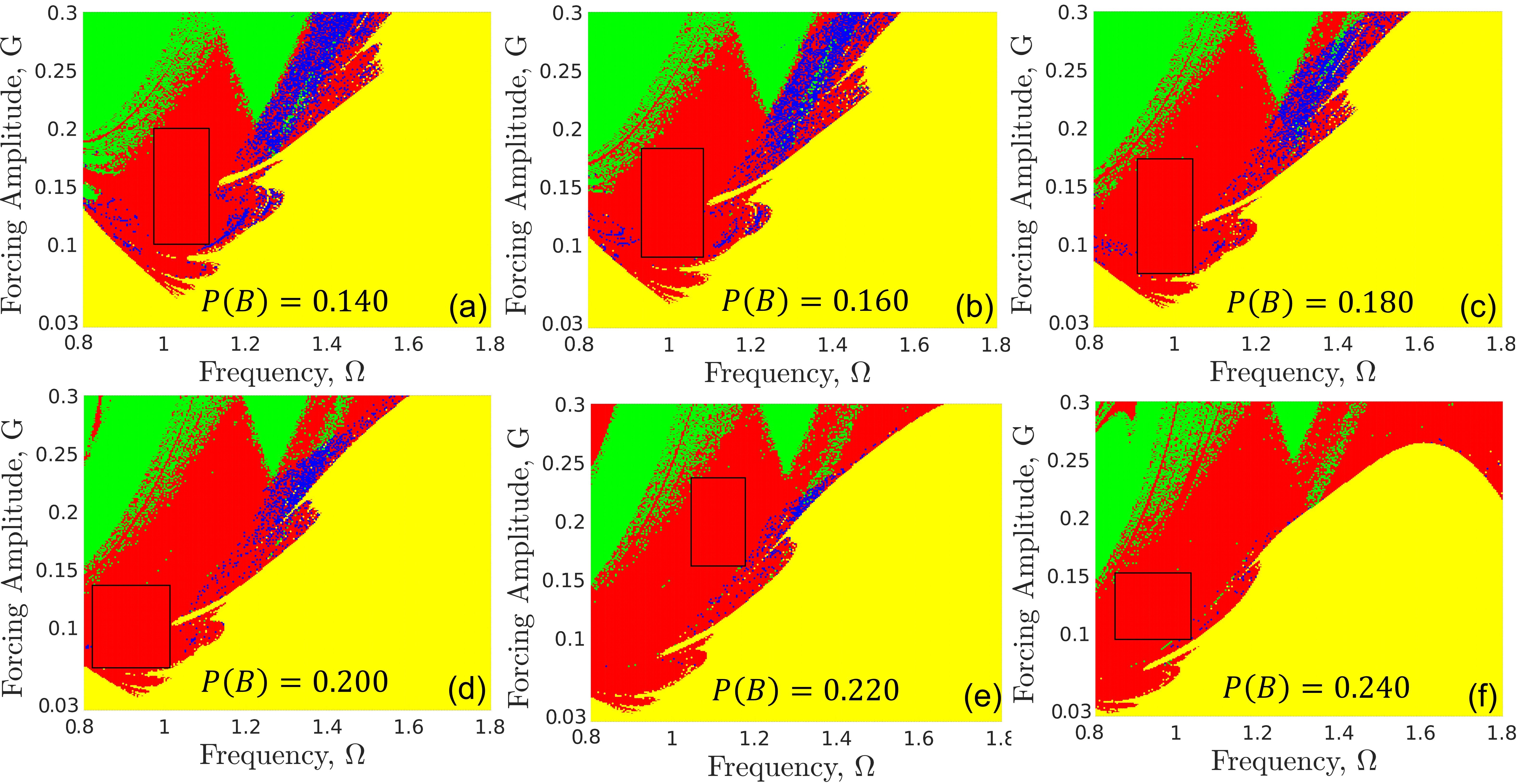}
\caption{Forcing amplitude-frequency parameter space for a bistable buckled beam with a damping ratio of \(\gamma = 0.07\) across static bias forces from \(P\textrm{(B)} = 0.140\) to \(0.240\) in increments of \(0.02\). A distinct switching area is marked by a rectangle indicating where every combination of \(G\) and \(\Omega\) results in switching behavior.}
\label{fig:F6}
\end{figure}

%%%%%%% Start of Figure 7 %%%%%%%
\begin{figure}[htbp]
\centering
\includegraphics[width=\textwidth]{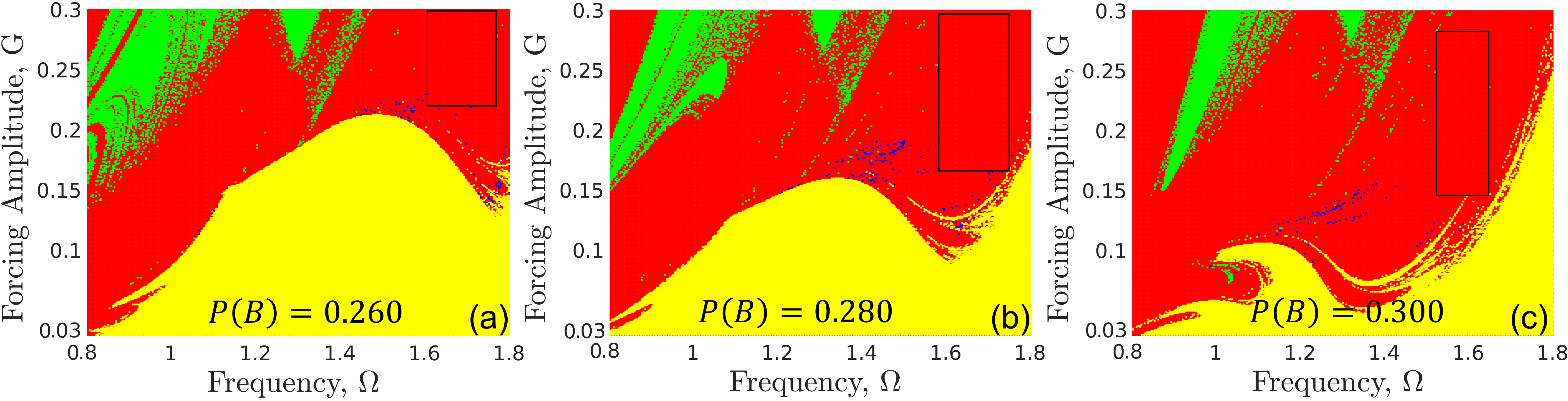}
\caption{Forcing amplitude-frequency parameter space for a bistable buckled beam with a damping ratio of \(\gamma = 0.07\) across static bias forces from \(P\textrm{(B)} = 0.240\) to \(0.300\) in increments of \(0.02\). A distinct switching area is marked by a rectangle indicating where every combination of \(G\) and \(\Omega\) results in switching behavior.}
\label{fig:F7}
\end{figure}

%In our analysis of bistable beam dynamics, pivotal to the understanding of mechanical metamaterials and MEMS devices, we employ a shift transformation \( u_1 = q_1 - 1 \) to streamline the mathematical framework. This transformation, elegantly shifting the reference point of the system's state variable \( q_1 \), redefines the equilibrium to a normalized zero baseline, thus substantially simplifying the differential equations governing the system's behavior. This methodological refinement not only aids in linearizing the equations around the buckled state but also provides a normalized scale for the variable, enhancing interpretability of the system's response. The resultant simplification of the governing equations is particularly beneficial, reducing analytical complexity and facilitating computational efficiency—a method in resonance with the ethos of perturbative and asymptotic analytical techniques.
\clearpage
\section{Relationship between static bias force and the minimum forcing amplitude \(G_\textrm{min}\) of the parameter space}

%%%%%%% Start of Figure 7 %%%%%%%
\begin{figure}[htbp]
\centering
\includegraphics[width=0.45\textwidth]{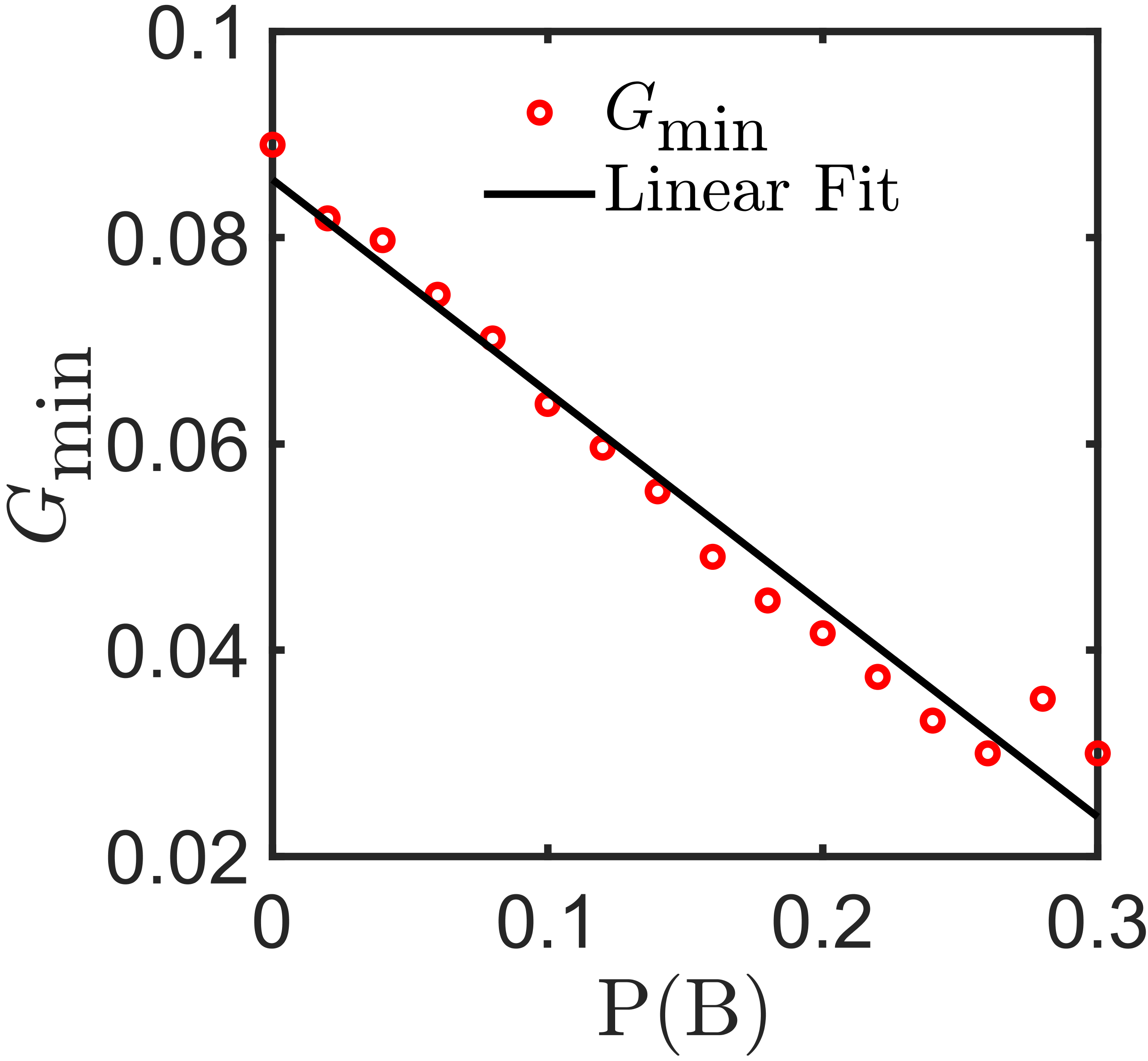}
\caption{Linear relationship between \(G_\textrm{min}\) and the applied bias force \(P\textrm{(B)}\). \(G_\textrm{min}\) is the minimum forcing amplitude required to switch between stable states in the parameter space shown in Figs. S2 to S4.}
\label{fig:8}
\end{figure}

\nocite{*}
%\bibliography{ref_SI}% Produces the bibliography via BibTeX.